\documentclass[twocolumn, journal]{IEEEtran}
\IEEEoverridecommandlockouts

\usepackage{color}
\usepackage{xcolor}
\usepackage{bm}
\usepackage{graphicx}
\usepackage{amsmath}
\usepackage{amssymb}
\usepackage{multirow}
\usepackage{algorithm}
\usepackage{algorithmic}
\usepackage{booktabs}
\usepackage{array}
\usepackage{amsthm}
\usepackage{lipsum}
\usepackage{enumerate}
\usepackage{stfloats}
\usepackage{subfigure}
\usepackage{cases}
\usepackage{cite}
\usepackage{mathtools}

\def\C{\mathbb{C}}

\def\Y{\mathbf{Y}}

\allowdisplaybreaks

\title{Lossy Beyond Diagonal Reconfigurable Intelligent Surfaces: Modeling and Optimization}

\author{Yiyang Peng, Hongyu Li,~\IEEEmembership{Member, IEEE}, Zheyu Wu, and Bruno Clerckx,~\IEEEmembership{Fellow, IEEE}
\thanks{Manuscript received 28 April 2025; revised 28 August 2025; accepted 31 October 2025. The associate editor coordinating the review
of this article and approving it for publication was Dr. Miao Wang. This work is supported by the National Natural Science Foundation of China (grant no. 62501509), and partially supported by UKRI grant EP/Y004086/1, EP/X040569/1, EP/Y037197/1, EP/X04047X/1, EP/Y037243/1.  \textit{(Corresponding author: Bruno Clerckx.)}}
\thanks{Yiyang Peng and Zheyu Wu are with the Department of Electrical and Electronic Engineering, Imperial College London, London SW7 2AZ, U.K. (e-mail:~yiyang.peng22@imperial.ac.uk;~zheyu.wu@imperial.ac.uk).}
\thanks{Hongyu Li is with the Internet of Things Thrust, The Hong Kong University of Science and Technology (Guangzhou), Guangzhou, Guangdong 511400, China (e-mail: hongyuli@hkust-gz.edu.cn).}
\thanks{Bruno Clerckx is with the Department of Electrical and Electronic Engineering, Imperial College London, London SW7 2AZ, U.K., and also with Kyung Hee University, Seoul, Korea (e-mail:~b.clerckx@imperial.ac.uk).}
}

\begin{document}
\maketitle
\thispagestyle{empty}
\begin{abstract}
    Beyond diagonal reconfigurable intelligent surface (BD-RIS) has emerged as an advancement and generalization of the conventional diagonal RIS (D-RIS) by introducing tunable interconnections between RIS elements, enabling smarter wave manipulation and enlarged coverage. While BD-RIS has demonstrated advantages over D-RIS in various aspects, most existing works rely on the assumption of a lossless model, leaving practical considerations unaddressed. This paper thus proposes a lossy BD-RIS model and develops corresponding optimization algorithms for various BD-RIS-aided communication systems. First, by leveraging admittance parameter analysis, we model each tunable admittance component based on a lumped circuit with losses and derive an expression of a circle characterizing the real and imaginary parts of each tunable admittance. We then consider the received signal power maximization in single-user single-input single-output (SISO) systems with the proposed lossy BD-RIS model. To solve the formulated challenging optimization problem, we design an effective algorithm by carefully exploiting the problem structure. In particular, an alternating direction method of multipliers (ADMM) framework is custom-designed to deal with the complicated constraints associated with lossy BD-RIS. Furthermore, we extend the proposed algorithmic framework to more general multiuser multiple-input single-output (MU-MISO) systems, where the transmit precoder and BD-RIS scattering matrix are jointly designed to maximize the sum-rate of the system. Finally, simulation results demonstrate that all BD-RIS architectures still outperform D-RIS in the presence of losses, but the optimal BD-RIS architectures in the lossless case are not necessarily optimal in the lossy case, e.g., group-connected BD-RIS can outperform fully- and tree-connected BD-RISs in SISO systems with relatively high losses at BD-RIS, whereas the opposite always holds true in the lossless case.
\end{abstract}


\section{Introduction}
Reconfigurable intelligent surface (RIS) is a nearly-passive planar surface composed of numerous reflecting elements, each capable of inducing a phase and/or amplitude change to the impinging wave. This flexibility enables dynamic wave manipulation, which provides enhanced channel gain and enlarges wireless coverage \cite{smart_radio_Marco_Di_renzo,IRS_magazine_Qingqing_Wu,IRS_tutorial_Qingqing_Wu}. According to microwave network theory \cite{pozar2009microwave}, RIS can be generally modeled as numerous scattering elements connected to a multiport reconfigurable impedance network \cite{shen2021modelling}. Most existing studies focus on RIS designs with a single-connected architecture, where each port of the reconfigurable impedance network is connected to the ground via a tunable impedance. This mathematically results in a diagonal scattering matrix \cite{shen2021modelling}, and such a single-connected RIS is also referred to as diagonal (D)-RIS. Recently, a novel branch of RIS, known as beyond diagonal (BD)-RIS \cite{BD-RIS_magazine}, has been proposed as a generalization of D-RIS. BD-RIS operates by introducing additional tunable impedance components that interconnect the ports in the reconfigurable impedance network, mathematically resulting in a scattering matrix that is no longer restricted to being diagonal. This enhanced flexibility, provided by those interconnections, enables a smarter wave manipulation, leading to greater channel gain, higher spectral efficiency, and broader wireless coverage compared to D-RIS \cite{BD-RIS_magazine}.

Existing studies on BD-RIS have shown its advantages over D-RIS in various aspects. These works can be categorized as follows: \cite{ios_hongliang_zhang,hybrid_hongyu,multisector_hongyu} focus on mode analysis; \cite{shen2021modelling,nerinigraphtheory,dynamic_grouping_hongyu,band_stem_zheyu,non_reciprocal_hongyu} focus on architecture design; \cite{close_form_nerini,channel_estimation_hongyu,universal_optimization_zheyu} focus on signal processing; \cite{discrete_nerini,mutual_coupling_hongyu,mutual_coupling_nerini,lossy_interconnections_nerini,wideband_circuit_hongyu} focus on hardware impairments. However, most existing works \cite{shen2021modelling,ios_hongliang_zhang,hybrid_hongyu,multisector_hongyu,nerinigraphtheory,dynamic_grouping_hongyu,close_form_nerini,channel_estimation_hongyu,non_reciprocal_hongyu,universal_optimization_zheyu,discrete_nerini,mutual_coupling_hongyu,mutual_coupling_nerini,wideband_circuit_hongyu,band_stem_zheyu} assume the reconfigurable impedance network is lossless to simplify the theoretical analysis and explore the performance upperbound of BD-RIS architectures. One exception work \cite{lossy_interconnections_nerini} models BD-RIS with lossy interconnections in the reconfigurable impedance network using transmission line theory, where these interconnections are in series with tunable impedance components. However, the tunable impedance components themselves are still assumed to be lossless. In practice, semiconductor devices such as positive-intrinsic-negative (PIN) diodes and varactor diodes are used to implement the tunable impedance components. These semiconductor devices have inherent resistances, which dissipate power in the circuit. Therefore, assuming a reconfigurable impedance network with lossless tunable components results in a model that underestimates the performance degradation in realistic systems. It is crucial to analyze the characteristics of BD-RIS with lossy components in the reconfigurable impedance network and establish an accurate model that reflects these losses.

In the works on practical D-RIS modeling \cite{practicalRIS_Abeywickrama,practical_ris_wenhao,practical_mimo_ofdm_ris_hongyu}, a lumped circuit model for each RIS element has been shown to effectively capture the relationship between reflection amplitude and phase. However, generalizing this approach to lossy BD-RIS is challenging. D-RISs are typically modeled using scattering parameters, where each entry in the scattering matrix represents the amplitude and phase response of an individual RIS element, allowing them to be modeled independently. In contrast, for BD-RISs, the losses of BD-RIS are embedded in each lossy reconfigurable component, and the interconnections between RIS elements cause the amplitude and phase of each entry in the scattering matrix to depend on other entries as well. In this sense, it is impossible to model lossy BD-RIS by independently characterizing its entries in the scattering matrix. Consequently, while lossless BD-RIS can be modeled directly by a beyond-diagonal scattering matrix, which is an upgrade of lossless D-RIS, the lossy BD-RIS needs an accurate model to reflect the losses embedded in the circuit, which, unfortunately, cannot be directly shown in the scattering matrix. In \cite{universal_framework_nerini}, a universal framework has been proposed that unifies the analysis of scattering, impedance, and admittance parameters for characterizing RIS-aided models, and presents the advantages of each type of parameter analysis in RIS modeling and optimization. While the aforementioned works \cite{shen2021modelling,discrete_nerini,close_form_nerini,hybrid_hongyu,multisector_hongyu,nerinigraphtheory,dynamic_grouping_hongyu,channel_estimation_hongyu,non_reciprocal_hongyu} primarily focus on modeling and designing BD-RIS-aided systems using scattering parameter, admittance parameter has proven to be a powerful approach in characterizing the circuit topology models of BD-RIS \cite{lossy_interconnections_nerini,wideband_circuit_hongyu,universal_optimization_zheyu}. Admittance parameter analysis enables the modeling of BD-RIS with lossy interconnections using transmission line theory \cite{lossy_interconnections_nerini}, the wideband modeling of BD-RIS through circuit-based approaches \cite{wideband_circuit_hongyu}, and the development of an efficient architecture-independent optimization framework for BD-RIS in general multiuser systems \cite{universal_optimization_zheyu}.

Inspired by the advantages of admittance parameter analysis as in \cite{lossy_interconnections_nerini,wideband_circuit_hongyu,universal_optimization_zheyu} to model and optimize BD-RIS, we propose to model lossy BD-RIS by individually accounting for each lossy tunable component by its admittance parameter. The contributions of this work are summarized as follows.

\textit{First}, we propose a novel model for BD-RIS that incorporates losses in the reconfigurable admittance network. Specifically, by utilizing admittance parameter analysis, we represent the lossy reconfigurable admittance network by modeling each tunable admittance component individually using a lumped circuit model based on a practical varactor. In the proposed model, the real and imaginary parts of each admittance component form a circle, with feasible values constrained to a segment of the circle determined by the practical range of the tunable capacitance. The proposed model is sufficiently general to be applicable to any BD-RIS architectures, such as fully-/group-connected \cite{shen2021modelling} and tree-/forest-connected architectures\footnote{The proposed model can also be applied to band-connected and stem-connected architectures \cite{band_stem_zheyu}.} \cite{nerinigraphtheory}.

\textit{Second}, we optimize the scattering matrix of BD-RIS to maximize the received signal power for a BD-RIS-aided single-user single-input single-output (SISO) system, incorporating the proposed lossy BD-RIS model. To solve this problem, we employ a minorization-maximization (MM) framework to transform the problem into a more tractable form. To deal with the complicated constraints introduced by lossy BD-RIS, we further custom-design an alternating direction method of multipliers (ADMM) algorithm to solve the resulting subproblems. In addition, to reduce the computational complexity, we propose a low-complexity solution that minimizes the Euclidean distance of an admittance matrix condition derived from the lossless BD-RIS model, using a similar ADMM algorithm to handle the constraints introduced by lossy BD-RIS.

\textit{Third}, we formulate the sum-rate maximization problem for a BD-RIS-aided multiuser multiple-input single-output (MU-MISO) system, jointly optimizing the transmit precoder and scattering matrix while incorporating the proposed lossy BD-RIS model. To tackle this problem, we employ the fractional programming (FP) technique to simplify the sum-rate expression and adopt a block coordinate descent (BCD) framework to solve the reformulated problem. In particular, an ADMM algorithm similar to the single-user case is applied to update the scattering matrix.

\textit{Fourth}, we present simulation results to evaluate the performance of the proposed algorithms. Results show that all BD-RIS architectures outperform D-RIS under the lossy RIS modeling. In contrast to the lossless case where fully- and tree-connected BD-RISs always outperform group-connected BD-RIS in SISO systems, in the lossy case, group-connected BD-RIS can outperform fully- and tree-connected BD-RISs. Except for the fully-connected architecture, group-connected BD-RIS consistently outperforms forest-connected BD-RIS under the same group size\footnote{Recall that in lossless BD-RIS, forest-connected and group-connected architectures achieve the same performance for a given group size in SISO systems \cite{nerinigraphtheory,parteo_frontier_nerini}, though the latter has higher complexity.}. In MU-MISO systems, increasing interconnections does not always lead into better performance as well. Consequently, the Pareto frontier characterized for lossless BD-RIS \cite{parteo_frontier_nerini,band_stem_zheyu} does not hold in the lossy case.

\textit{Organization:} 
Section \ref{sec:model_lossy} introduces a lossy BD-RIS model based on admittance parameter analysis. Section \ref{sec:siso} presents the system model and problem formulation for BD-RIS-aided SISO systems using the proposed lossy model and proposes two iterative algorithms to solve the problem. Section \ref{sec:mu_miso} extends the system model and problem formulation to BD-RIS-aided MU-MISO systems and proposes an iterative algorithm to solve it. Section \ref{sec:simulation} provides simulations to evaluate the performance of the proposed lossy BD-RIS model and algorithms. Finally, Section \ref{sec:conclusion} concludes the paper.

\textit{Notations:}
Boldface lowercase and uppercase letters represent column vectors and matrices, respectively, while scalars are denoted with letters not in bold font. 
$\mathbb{R}$ and $\mathbb{C}$ indicate the set of real numbers and complex numbers, respectively. 
$(\cdot)^*$, $(\cdot)^T$, $(\cdot)^H$, and $(\cdot)^{-1}$ represent the conjugate, transpose, conjugate transpose, and inverse, respectively. 
$\Re\{\cdot\}$ and $\Im\{\cdot\}$ indicate the real and imaginary parts of a complex number, respectively. 
$\mathbb{E}\{\cdot\}$ denotes the statistical expectation. 
$|\cdot|$, $\| \cdot \|_2$ and $\| \cdot \|_F$ refer to the absolute-value norm, the $\ell_2$ norm, and the Frobenius norm, respectively. $\arg(\cdot)$ denotes the phase of a complex number. $\otimes$ represents the Kronecker product. 
$\mathsf{blkdiag}(\cdot)$ represents a block-diagonal matrix. 
$\mathsf{vec}(\cdot)$ denotes the vectorization of a matrix, $\overline{\mathsf{vec}}(\cdot)$ is the reverse operation of vectorization, and $\mathsf{tr}(\cdot)$ denotes the trace of a matrix.
$\mathbf{I}_M$ denotes an $M \times M$ identity matrix. 
$\jmath = \sqrt{-1}$ represents the imaginary unit.
$[\mathbf{A}]_{i,j}$ and $[\mathbf{a}]_i$ indicate the $(i,j)$-th entry of $\mathbf{A}$ and the $i$-th entry of $\mathbf{a}$, respectively.
Finally, $[\mathbf{A}]_{i:i',j:j'}$ represents the submatrix of $\mathbf{A}$ formed by selecting the $i$-th to $i'$-th rows and $j$-th to $j'$-th columns.

\section{Modeling of Lossy BD-RIS}\label{sec:model_lossy}
In this section, we first model BD-RIS using admittance parameters and then characterize the loss of BD-RIS based on a lumped circuit model.

\subsection{Admittance Parameter Analysis of BD-RIS}\label{sec:Y_parameter}
An $M$-element RIS is a passive device, which is modeled as $M$ elements connected to an $M$-port reconfigurable admittance network \cite{shen2021modelling}. The reconfigurable admittance network, composed of tunable passive components, can be mathematically described by its admittance matrix $\mathbf{Y}\in\mathbb{C}^{M\times M}$, or equivalently, its scattering matrix $\mathbf{\Phi}\in\mathbb{C}^{M\times M}$ \cite{universal_framework_nerini}. Specifically, the former characterizes the relationship between voltages and currents at the $M$ ports of the reconfigurable admittance network, while the latter characterizes the scattering behavior relating the incident and reflected waves. According to microwave network theory \cite{pozar2009microwave}, we can map the scattering matrix $\bm \Phi$ to the admittance matrix $\mathbf{Y}$ by
\begin{equation}\label{eq:Phi_Y}
    \bm \Phi = \left( Y_0\mathbf{I}_M + \mathbf{Y}\right)^{-1} \left( Y_0\mathbf{I}_M - \mathbf{Y} \right),
\end{equation}
where $Y_0$ is the characteristic admittance and usually set as $Y_0 = \frac{1}{50}$ siemens (S). Depending on the circuit topology of the $M$-port reconfigurable admittance network, the scattering matrix $\bm \Phi$ and admittance matrix $\mathbf{Y}$ exhibit different mathematical characteristics. In this paper, we consider two representative architectures of BD-RIS: the group-connected architecture \cite{shen2021modelling} and the forest-connected architecture \cite{nerinigraphtheory}. The detailed descriptions of these two architectures are shown below.

\subsubsection{Group-Connected Reconfigurable Admittance Network}
In this architecture, the $M$ ports of the reconfigurable admittance network is uniformly divided into $G$ groups. The ports within the same group are all connected to each other via tunable admittance components, whereas the ports in different group are not connected to each other \cite{shen2021modelling}. Fig. \ref{fig:1}(a) shows a 36-element BD-RIS with the group-connected architecture of group size 3, where one group is randomly selected for illustration. For a group-connected reconfigurable admittance network with $G$ uniform groups, each group contains $\Bar{M} = \frac{M}{G}$ elements, referred to as the group size. For the $g$-th group, all ports are connected with each other, leading to a full admittance matrix $\mathbf{Y}_g\in\mathbb{C}^{\bar{M}\times\bar{M}}$. Hence, the admittance matrix $\mathbf{Y}$ has a block-diagonal structure given by
\begin{equation}\label{eq:blkdiag_Y}
    \mathbf{Y} = \mathsf{blkdiag} \left(\Y_1,\Y_2,\ldots,\Y_G\right).
\end{equation}
For a reciprocal reconfigurable admittance network, $\Y_g,~\forall g \in \mathcal{G} = \{1,\ldots,G\}$ is symmetric, that is 
\begin{equation}\label{eq:symmetric_Y}
    \Y_g = \Y^T_g, ~\forall g \in \mathcal{G}.
\end{equation}
More specifically, as illustrated in Fig. \ref{fig:1}(a), in the $g$-th group of the reconfigurable admittance network, each port $m_g = (g-1)\bar{M}+m$ is connected to port $n_g = (g-1)\bar{M}+n$ through a tunable admittance $Y_{m_g,n_g}, ~\forall m,n \in \bar{\mathcal{M}}= \{1,\ldots,\bar{M}\}$. In addition, each port $m_g$ is also connected to ground via an admittance $Y_{m_g,m_g}$. According to \cite{pozar2009microwave}, $\Y_g$ is written as
\begin{equation}
    [\mathbf{Y}_g]_{m,n} = \begin{cases}
        -Y_{m_g,n_g}, &m_g\ne n_g; \\
        \sum_{k=(g-1)\bar{M}+1}^{g\bar{M}}Y_{m_g,k}, & m_g = n_g.
    \end{cases}
    \label{eq:Y_calculate}
\end{equation}
The circuit complexity of group-connected BD-RIS, defined as the required number of tunable admittance components, is given by $\frac{M(\bar{M}+1)}{2}$. An extreme case is the fully-connected architecture \cite{shen2021modelling}, with $\bar{M} = M$, and thus a circuit complexity of $\frac{M(M+1)}{2}$. The circuit complexity of the group-connected architecture can be significantly high, particularly when the group size $\bar{M}$ is large. To reduce this complexity while maintaining the flexibility of the reconfigurable admittance network, the forest-connected architecture for BD-RIS has been proposed in \cite{nerinigraphtheory}.

\subsubsection{Forest-Connected Reconfigurable Admittance Network}
\begin{figure}
    \centering
    \includegraphics[width=0.47\textwidth]{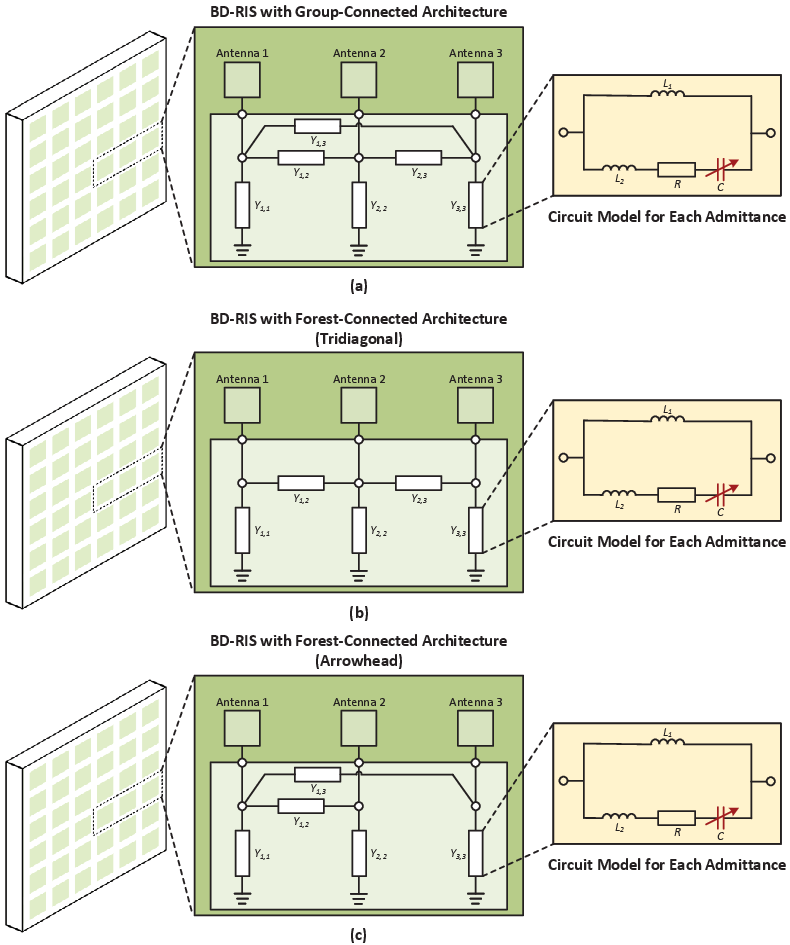}
    \caption{Examples of a 36-element BD-RIS with (a) a group-connected reconfigurable admittance network, (b) a forest-connected reconfigurable admittance network in tridiagonal form, and (c) a forest-connected reconfigurable admittance network in arrowhead form, each with a group size of 3, along with the corresponding circuit model for each admittance component.}\label{fig:1}
\end{figure}

In a forest-connected BD-RIS with $G$ uniform groups, each group $g$ follows a tree-connected architecture \cite{nerinigraphtheory}, which means each port is only directly connected to one other port via a tunable admittance component. This tree-connected architecture can take the form of either a tridiagonal RIS or an arrowhead RIS. Two 36-element forest-connected BD-RISs with a group size 3 in tridiagonal form and arrowhead form are illustrated in Fig. \ref{fig:1}(b) and Fig. \ref{fig:1}(c), respectively. In a tridiagonal BD-RIS, every pair of adjacent ports within the $\bar{M}$ ports is connected via a tunable admittance component. In an arrowhead BD-RIS, one designated port among the $\bar{M}$ ports connects to all other ports, while the remaining ports are connected only to the ground. Both tree-connected forms have the same circuit complexity. In this case, the circuit complexity of a forest-connected BD-RIS reduces to $(2\bar{M}-1)G$. The admittance matrix $\Y$ still satisfies (\ref{eq:blkdiag_Y}) -- (\ref{eq:Y_calculate}). In addition, for the tridiagonal form, the following constraint is imposed:
\begin{equation}\label{eq:Y_tri}
    Y_{m_g,n_g}=0, ~\forall m_g,n_g, ~|m_g-n_g| > 1.
\end{equation}
For the arrowhead form, where the designated port $c_g$ connects all other ports, the following constraint is added:
\begin{equation}\label{eq:Y_arr}
    Y_{m_g, n_g} = 0,~\forall m_g, n_g, ~m_g \neq c_g, ~n_g \neq c_g, ~m_g \neq n_g.
\end{equation}

Based on the above illustrations, the D-RIS with a single-connected architecture where each port is connected to ground via its own admittance component without interacting with other ports is a special case of BD-RIS with both group and forest-connected architectures when $\bar{M}=1$. The simple architecture of D-RIS makes it possible to individually model the loss of each tunable component and reflect it directly to its diagonal scattering matrix $\bm \Phi$ \cite{practicalRIS_Abeywickrama}. However, this does not hold for BD-RIS, whose scattering matrix has coupled entries due to the interconnections. To tackle this issue, we use admittance parameters to facilitate the modeling of lossy BD-RIS. Applying admittance parameter allows us to establish a linear mapping between the tunable admittance components expressed by the circuit model and the admittance matrix $\mathbf{Y}$. This further benefits the analysis of the BD-RIS loss, which fundamentally comes from the resistance component in the circuit model. Consequently, modeling the scattering matrix can be converted into modeling each tunable admittance component. 

\subsection{Modeling of Lossy Tunable Admittance}\label{sec:model_lossy_admittance}
In practice, each tunable admittance component is realized using a semiconductor device that can adjust its admittance. Given that the physical length of a reconfigurable admittance component is usually much smaller than the wavelength of the incident signal, the response of each tunable admittance component can be accurately characterized by an equivalent lumped circuit model, regardless of its geometry realization \cite{koziel2013surrogate}. As illustrated in Fig. \ref{fig:1}, each tunable admittance $Y_{m_g,n_g}$ is modeled as a parallel resonant circuit\footnote{This circuit models a practical varactor diode, where the varactor is represented by a tunable capacitance $C$ in series with a resistor $R$ and an inductor $L_2$, along with a parallel inductor $L_1$. The resistor $R$ accounts for the parasitic resistance of the varactor, which includes the energy loss, while $L_2$ represents the parasitic inductance. From a practical design perspective, $L_1$ is incorporated to extend the tuning range of the circuit’s resonances. From an optimization perspective, $L_1$ is introduced to adjust the feasible susceptance values of each admittance component.} \cite{koziel2013surrogate}. This lumped circuit model, initially used in D-RIS to capture the relationship between reflection amplitude and phase shift for each reflecting element \cite{practicalRIS_Abeywickrama}, is generalized in BD-RIS to model the losses of each admittance component. The admittance of this equivalent circuit is given by
\begin{equation}
    Y_{m_g,n_g}(C_{m_g,n_g}) = \frac{1}{\jmath\omega L_1} + \frac{1}{\jmath\omega L_2 +\frac{1}{\jmath \omega C_{m_g,n_g}} + R}, ~\forall m_g,n_g,\label{eq:admittance}
\end{equation}
where $\omega = 2\pi f$ represents the angular frequency of the circuit. In particular, the loss of BD-RIS is characterized by the resistor $R$, which determines the amount of power dissipation in the reconfigurable admittance network and can not be ignored in the implementation of practical semiconductor devices. The values of $L_1$, $L_2$, and $R$ are fixed and the same for each admittance component, while the admittance of the entire circuit model is tuned by adjusting the capacitance $C_{m_g,n_g}$. 

\begin{figure}
    \centering
    \includegraphics[width=0.5\textwidth]{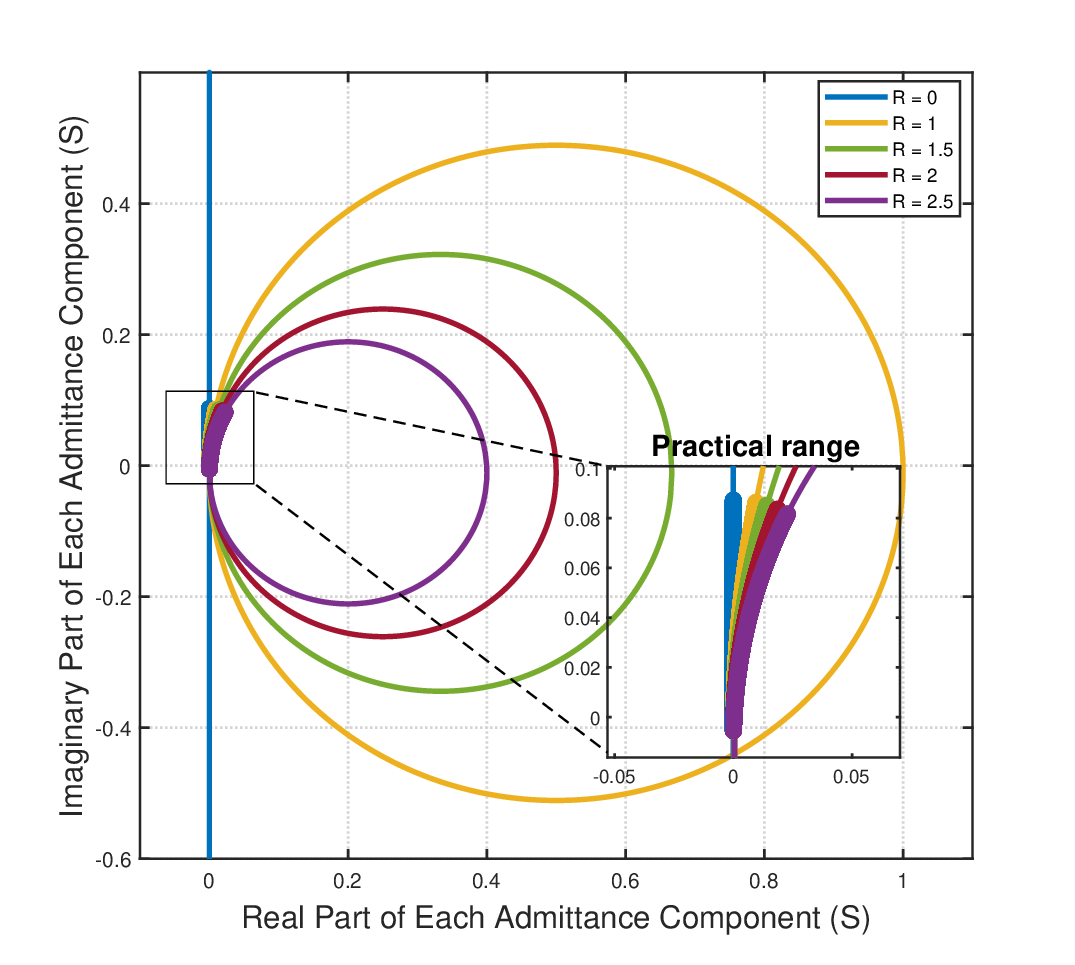}
    \caption{Imaginary part of each $Y_{m_g,n_g}$ as a function of its real part at a signal frequency of $f = 2.4$ GHz with $L_1 = 6$ nH. Circles represent all possible values of the admittance component for varying $R$. The practical range corresponds to $C_{m_g,n_g} \in [0.35, 3.20]$ pF and $L_2 = 0.7$ nH.}\label{fig:2}
\end{figure}

To obtain more insights into how the value of the admittance component $Y_{m_g,n_g}$ varies with capacitance $C_{m_g,n_g}$, we re-write expression (\ref{eq:admittance}) to explicitly separate and its real and imaginary parts as\begin{equation}
    \begin{aligned}
        Y_{m_g,n_g}(&C_{m_g,n_g}) = \frac{R}{R^2 + \left(\omega L_2 - \frac{1}{\omega C_{m_g,n_g}} \right)^2}  \\
    &+ \jmath \left( -\frac{1}{\omega L_1} + \frac{-\omega L_2 + \frac{1}{\omega C_{m_g,n_g}}}{R^2 + \left(\omega L_2 - \frac{1}{\omega C_{m_g,n_g}} \right)^2} \right). \label{eq:admittance_simplified}
    \end{aligned}
\end{equation}
An interesting observation drawn from (\ref{eq:admittance_simplified}) is that the real and imaginary parts of each tunable admittance component describe a circle in the complex plane, given by
\begin{equation}
    \begin{aligned}
        &\left(\Re\{ Y_{m_g,n_g} \} - \frac{1}{2R} \right)^2  \\
        &\quad + \left(\Im\{ Y_{m_g,n_g} \} - \left(-\frac{1}{\omega L_1} \right) \right)^2 = \left( \frac{1}{2R} \right)^2. \label{eq:circle}
    \end{aligned}
\end{equation}
To the best of our knowledge, the circular relationship in (\ref{eq:circle}) has not been reported in prior literature and is analytically derived in this paper for the first time. Clearly, the radius of this circle is $\frac{1}{2R}$, which depends solely on the resistance $R$, and the center of the circle is located at $\left(\frac{1}{2R}, -\frac{1}{\omega L_1} \right)$, which depends jointly on the resistance $R$ and the inductance $L_1$. In this paper, we utilize the parameters from the data sheet of the practical varactor diode SMV2020-079LF to model the equivalent circuit of each admittance component. Specifically, we set $L_2 = 0.7$ nH, $C_{m_g,n_g} \in [0.35, 3.20]$ pF\footnote{The limited range of capacitance values, along with the presence of $L_1$ prevents the tunable admittances of interconnections from being completely "turned off", indicating that single-connected architecture is not a subset of the fully-/group-connected and tree-/forest-connected architectures in BD-RIS.} and assume $L_1 = 6$ nH\footnote{In \cite{practicalRIS_Abeywickrama}, $L_1$ is assumed to be 2.5 nH to achieve the full reflection tuning range for each reflecting element. However, this assumption is unnecessary in our BD-RIS model due to the nonlinear mapping from the admittance matrix to scattering matrix. The selection of the $L_1$ value is discussed in Section \ref{sec:simulation}.}. As such, we represent the possible values of $Y_{m_g,n_g}$ through the set $\mathcal{Y}$, defined as
\begin{equation}\label{eq:Y_set}
    \mathcal{Y}= \{\alpha + \beta e^{\jmath \theta} \ \mid \ \theta \in [\theta_\mathrm{min}, \theta_\mathrm{max}]\},
\end{equation}
where $\alpha$ and $\beta$ are defined as $\alpha \triangleq \left( \frac{1}{2R} - \jmath \frac{1}{\omega L_1} \right)$ and $\beta \triangleq \frac{1}{2R}$, respectively, and $\theta$ is introduced to re-describe the complex circle in (\ref{eq:circle}) with $\theta_\mathrm{min}$ and $\theta_\mathrm{max}$ determined by the practical capacitance range of the varactor SMV2020-079. To visualize the constraint (\ref{eq:Y_set}), Fig. \ref{fig:2} illustrates four circles, each corresponding to a different resistance value $R$ ($R \neq 0$). Eq. (\ref{eq:circle}) indicates that the capacitance $C_{m_g,n_g}$ and inductance $L_2$ do not affect the shape of the circle. However, they do influence the specific values of the real and imaginary parts of $Y_{m_g,n_g}$. As shown in Fig. \ref{fig:2}, with a fixed $L_2$ and a limited range of $C_{m_g,n_g}$, only a segment of the circle is feasible, restricting the possible values of $Y_{m_g,n_g}$. 

\textit{Remark 1:} In most existing D-RIS \cite{smart_radio_Marco_Di_renzo,IRS_magazine_Qingqing_Wu,IRS_tutorial_Qingqing_Wu,cascaded_CE_ris_Guo,weighted_sum_rate_ris_Guo} and most existing BD-RIS literature\footnote{\cite{lossy_interconnections_nerini} studies lossy interconnections between admittance components while all the admittance components themselves are assumed to be lossless. The optimization methods proposed in \cite{lossy_interconnections_nerini} are not applicable to our work, as we focus on BD-RIS with lossy tunable admittance components.} \cite{shen2021modelling,BD-RIS_magazine,discrete_nerini,close_form_nerini,hybrid_hongyu,multisector_hongyu,nerinigraphtheory,channel_estimation_hongyu,non_reciprocal_hongyu,universal_framework_nerini,mutual_coupling_hongyu,mutual_coupling_nerini,wideband_circuit_hongyu,universal_optimization_zheyu,dynamic_grouping_hongyu,band_stem_zheyu,parteo_frontier_nerini}, the reconfigurable admittance network is assumed to be lossless, such that each tunable admittance is purely imaginary with $\Re\{Y_{m_g,n_g}\} = 0$ and the feasible range of its imaginary part lies unconstrained along the y-axis as illustrated in Fig. \ref{fig:2}. This simple assumption fails to capture the coupling between the real and imaginary parts of $Y_{m_g,n_g}$, which is important in practical realizations of BD-RIS and will have a non-negligible impact on the system performance as it will be illustrated in the following sections. 
 
\section{Optimization for BD-RIS-Aided SISO System}\label{sec:siso}
In this section, we apply the lossy BD-RIS modeling in SISO systems and formulate the problem of maximizing the received signal power. We propose two algorithms to optimize BD-RIS matrices.

\subsection{System Model and Problem Formulation}
We start by considering an $M$-element BD-RIS-aided SISO wireless communication system. In this paper, we adopt the common assumptions from related literature \cite{shen2021modelling,BD-RIS_magazine,discrete_nerini,close_form_nerini,hybrid_hongyu,multisector_hongyu,nerinigraphtheory,channel_estimation_hongyu,lossy_interconnections_nerini,wideband_circuit_hongyu,universal_optimization_zheyu,dynamic_grouping_hongyu,band_stem_zheyu,parteo_frontier_nerini}, which state that the antennas at the transmitter, RIS, and receiver are perfectly matched, have no mutual coupling and no structural scattering. The overall channel $h$ between the transmitter and receiver can thus be written as
\begin{equation}
    h = h_{RT} + \mathbf{h}_{RI}^H \bm \Phi \mathbf{h}_{IT}, \label{eq:channel_scattering}
\end{equation}
where $h_{RT} \in \C$, $\mathbf{h}_{RI} \in \C^{M \times 1}$ and $\mathbf{h}_{IT} \in \C^{M \times 1}$ denote the channels from the transmitter to receiver, from BD-RIS to receiver and from transmitter to BD-RIS, respectively. 

Next, we aim to maximize the received signal power for the considered BD-RIS-aided SISO system, taking into account the losses in BD-RIS. As discussed in Section \ref{sec:Y_parameter}, it is convenient to capture the losses of BD-RIS with admittance parameter. We introduce the following notation to define the BD-RIS architecture:
$$
\begin{aligned}\mathcal{I}_g=\big\{(m,n)\mid &\text{ the } m \text{-th and }n\text{-th} \text{ elements in the } g\text{-th group}\\&\text{ are not connected}\big\}.
\end{aligned}$$ In particular, $\mathcal{I}_g=\emptyset$ for group-connected BD-RIS, $\mathcal{I}_g=\{(m,n)\mid |m-n|>1\}$ for tridiagonal forest-connected BD-RIS, and $\mathcal{I}_g=\{(m,n)\mid m, n\neq c_g, m\neq n\}$ for arrowhead forest-connected BD-RIS. By employing the relationship between the scattering matrix $\bm \Phi$ and the admittance matrix $\mathbf{Y}$ in (\ref{eq:Phi_Y}), the mapping between $\mathbf{Y}$ and the tunable admittances $Y_{m_g,n_g}, ~\forall m_g,n_g$ in (\ref{eq:blkdiag_Y}) -- (\ref{eq:Y_calculate}),  and the set of feasible values for each $Y_{m_g,n_g}$ in (\ref{eq:Y_set}), the power maximization problem can be formulated as follows\footnote{We include $\bm \Phi,\{\mathbf{Y}_g\}$ and $\{Y_{m_g,n_g}\}$ as optimization variables because they are treated as independent variables in the optimization process.}:
\begin{subequations}\label{eq:received_signal_power_max}
    \begin{align}
        \max_{\bm \Phi,\{\mathbf{Y}_g\},\atop \{Y_{m_g,n_g}\}}~&\left| h_{RT} + \mathbf{h}_{RI}^H\bm \Phi \mathbf{h}_{IT} \right|^2 \label{eq:objective_theta}\\
        \mathrm{s.t.}~~~~&\bm \Phi=\left(Y_0\mathbf{I}_M+\mathbf{Y}\right)^{-1}\left(Y_0\mathbf{I}_M-\mathbf{Y}\right), \label{eq:theta_Y}\\
        &\mathbf{Y} = \mathsf{blkdiag}(\mathbf{Y}_{1},\ldots,\mathbf{Y}_{G}), \label{eq:blkdiag}\\
        &\mathbf{Y}_g = \mathbf{Y}_g^T, ~\forall g \in \mathcal{G},\label{eq:symmetric}\\
        & [\mathbf{Y}_g]_{m,n} = \begin{cases}
        -Y_{m_g,n_g}, &m_g\ne n_g \\
        \sum_{k=(g-1)\bar{M}+1}^{g\bar{M}}Y_{m_g,k}, & m_g = n_g
        \end{cases},\label{eq:Y_bar_cal}\\
        & Y_{m_g,n_g}=0, ~\forall (m_g,n_g)\in \mathcal{I}_g,\notag\\
        &Y_{m_g,n_g} \in \mathcal{Y},~\forall  (m_g,n_g)\notin \mathcal{I}_g.\label{eq:Y_range}
    \end{align}
\end{subequations}
Problem (\ref{eq:received_signal_power_max}) is challenging to solve because constraint (\ref{eq:theta_Y}) introduces a nonlinear mapping between the admittance matrix and the scattering matrix, resulting from matrix inversion. Additionally, constraint (\ref{eq:Y_range}) is non-convex, which further complicates the optimization process. In the following subsections, we propose two solutions to maximize the received signal power with a lossy BD-RIS in the SISO systems. First, we propose an MM-ADMM algorithm in Section \ref{sec:MM-ADMM} to directly solve (\ref{eq:received_signal_power_max}). In Section \ref{sec:low_complex_admm}, we further propose a low-complexity algorithm that approximately maximizes the received signal power without directly solving (\ref{eq:received_signal_power_max}). 

\textit{Remark 2:} In lossless BD-RIS modeling, constraint (\ref{eq:Y_bar_cal}) is typically not considered because there are no restrictions on individual tunable components $Y_{m_g,n_g},~\forall m_g,n_g$. Furthermore, constraint (\ref{eq:Y_range}) arises specifically from modeling lossy tunable admittance components, which does not exist in lossless BD-RIS models. More specifically, take a lossless fully-connected BD-RIS as an example, each $Y_{m_g,n_g}$ has zero real part and arbitrary imaginary part, which leads to $\Y = \jmath \mathbf{B}$, with $\mathbf{B}$ being an arbitrary real-valued matrix. Consequently, the scattering matrix $\bm\Phi$ is an arbitrary unitary matrix. Therefore, one can focus on the design of either unconstrained $\mathbf{B}$ or constrained $\bm\Phi$ for lossless BD-RIS. However, neither of the above two strategies work for lossy BD-RIS since the real and imaginary parts of $Y_{m_g,n_g}$ is inherently coupled and constrained by (\ref{eq:Y_range}). These two constraints (\ref{eq:Y_bar_cal}) and (\ref{eq:Y_range}), together with the nonlinear relation in (\ref{eq:theta_Y}), constitute the main challenges introduced when accounting for lossy tunable components compared to lossless BD-RIS models.

\subsection{Proposed MM-ADMM Algorithm}\label{sec:MM-ADMM}
Defining $\Tilde{\mathbf{h}}_{RI} \triangleq h_{RT}\mathbf{h}_{RI}$, the objective function in (\ref{eq:objective_theta}) can be re-written as $\left| h_{RT} + \mathbf{h}_{RI}^H\bm \Phi \mathbf{h}_{IT} \right|^2= |h_{RT}|^2 + 2\Re\{\Tilde{\mathbf{h}}^H_{RI} \bm \Phi \mathbf{h}_{IT}\} + \left| \mathbf{h}_{RI}^H\bm \Phi \mathbf{h}_{IT} \right|^2$. Defining $\hat{\mathbf{H}}_{RI} \triangleq \mathbf{h}_{RI} \mathbf{h}_{RI}^H$ and $\hat{\mathbf{H}}_{IT} \triangleq \mathbf{h}_{IT} \mathbf{h}_{IT}^H$, we can re-write $\left| \mathbf{h}_{RI}^H\bm \Phi \mathbf{h}_{IT} \right|^2$ as
\begin{align}
    \left| \mathbf{h}_{RI}^H\bm \Phi \mathbf{h}_{IT} \right|^2 &=
    \mathsf{tr}(\mathbf{h}_{RI}\mathbf{h}_{RI}^H\bm \Phi\mathbf{h}_{IT}\mathbf{h}_{IT}^H\bm \Phi^H) \nonumber \\
    &= \mathsf{vec}^T(\bm \Phi^*) [\hat{\mathbf{H}}_{IT}^T \otimes \hat{\mathbf{H}}_{RI}] \mathsf{vec}(\bm \Phi) \nonumber \\
    &= \Tilde{\bm\phi}^H \mathbf{F}\Tilde{\bm\phi},\label{eq:signal_power_re}
\end{align}
where we define $\Tilde{\bm\phi} \triangleq \mathsf{vec}(\bm \Phi)$, $\bm \Phi = \mathsf{blkdiag}(\bm \Phi_{1},\ldots,\bm \Phi_{G})$ with $\bm \Phi_g \in \C^{\bar{M}\times\bar{M}}$, $\mathbf{F} \triangleq \hat{\mathbf{H}}_{IT}^T \otimes \hat{\mathbf{H}}_{RI} \in \C^{M^2 \times M^2}$ and utilize the property $\mathsf{tr}(\mathbf{A}\mathbf{B}\mathbf{C}\mathbf{D}) = \mathsf{vec}^T(\mathbf{D}^T) [\mathbf{C}^T \otimes \mathbf{A}] \mathsf{vec}(\mathbf{B})$. The term $\{\Tilde{\mathbf{h}}^H_{RI} \bm \Phi \mathbf{h}_{IT}\}$ can be reformulated using the vectorization property as
\begin{equation}
    \Tilde{\mathbf{h}}^H_{RI} \bm \Phi \mathbf{h}_{IT} = \boldsymbol{l}^H \Tilde{\bm\phi},
\end{equation}
where $\boldsymbol{l} \triangleq (\mathbf{h}_{IT}^T \otimes \Tilde{\mathbf{h}}^H_{RI})^H$. The original problem (\ref{eq:received_signal_power_max}) can thus be equivalently transformed into
\begin{subequations}\label{eq:phi_F_phi}
    \begin{align}
        \max_{\Tilde{\bm\phi}, \{\Y_g\}, \{Y_{m_g,n_g}\}}~~&\Tilde{\bm\phi}^H \mathbf{F}\Tilde{\bm\phi} + 2\Re\{\boldsymbol{l}^H \Tilde{\bm\phi}\}   \label{eq:phi_F_phi_a}\\
        \mathrm{s.t.}~~~~~~~~~ &\text{(\ref{eq:theta_Y}) -- (\ref{eq:Y_range}).}
    \end{align}
\end{subequations}
Let $\bar{\Y}_g \in \C^{\bar{M} \times \bar{M}}$ be a matrix with entries $[\bar{\Y}_g]_{m,n} = Y_{m_g,n_g}$, then (\ref{eq:Y_bar_cal}) can be expressed as a linear mapping from $\bar{\Y}_g$ to $\Y_g, ~\forall g \in \mathcal{G}$. To facilitate the optimization, we further re-write this linear mapping into a vector form as $\mathsf{vec}(\Y_g) = \mathbf{B} \mathsf{vec}(\bar{\Y}_g)$, where matrix $\mathbf{B} \in \{0,\pm 1\}^{\bar{M}^2 \times \bar{M}^2}$ maps $\mathsf{vec}(\bar{\Y}_g)$ to $\mathsf{vec}(\Y_g)$, $~\forall g \in \mathcal{G}$, and is defined as
\begin{equation}
     \begin{aligned}
        \relax[\mathbf{B}]_{\bar{M}(i-1)+j,l} = 
        \begin{cases}
            1, & l = \bar{M}(k-1) + i ~\text{and}~ j=i;\\
            -1, & l = \bar{M}(i-1) + j ~\text{and}~ j \neq i;\\
            0, & \text{otherwise},
        \end{cases}
    \end{aligned}
\end{equation}
$\forall i,j,k \in \mathcal{\bar{M}}$. It is worth noting that $\bar{\Y}_g, ~\forall g \in \mathcal{G}$, is symmetric due to the symmetry of $\mathbf{Y}_g$ and the relation between $\mathbf{Y}_g$ and $\bar{\Y}_g$ in (\ref{eq:Y_bar_cal}). In addition, the BD-RIS architecture determines the locations of non-zero elements in $\bar{\mathbf{Y}}_g$; see (\ref{eq:Y_range}). Hence, the actual degrees of freedom to be designed in $\bar{\mathbf{Y}}_g$ are the non-zero entries in its upper-triangular part, whose cardinality equals the number of tunable admittance components in each BD-RIS group, denoted by $U$. Let $\bar{\mathbf{y}}_g \in \C^{U \times 1}$ be the vector collecting these entries. The symmetry and architecture-related constraints can then be jointly incorporated by rewriting $\mathsf{vec}(\bar{\Y}_g)=\mathbf{P}\bar{\mathbf{y}}_g,~\forall g\in\mathcal{G}$, where $\mathbf{P} \in \{0,1\}^{\bar{M}^2 \times U}$ is a binary matrix mapping $\bar{\mathbf{y}}_g$ to $\mathsf{vec}(\bar{\Y}_g)$. We provide the explicit forms of matrix $\mathbf{P}$ and $U$ for the group-connected and forest-connected architectures in the Appendix.

According to the aforementioned discussions, problem (\ref{eq:phi_F_phi}) can be equivalently re-written as
\begin{subequations}\label{eq:phi_F_phi_reformulate}
    \begin{align}        \max_{\Tilde{\bm\phi},\{\mathbf{Y}_g\},\{\bar{\mathbf{y}}_g\}}~~&\Tilde{\bm\phi}^H \mathbf{F}\Tilde{\bm\phi} + 2\Re\{\boldsymbol{l}^H \Tilde{\bm\phi}\}\label{eq:phi_F_phi_re_a}\\  \mathrm{s.t.}~~~~~~&\bm\Phi_g=\left(Y_0\mathbf{I}_{\bar{M}}+\mathbf{Y}_g\right)^{-1}\left(Y_0\mathbf{I}_{\bar{M}}-\mathbf{Y}_g\right)
        ,~\forall g \in \mathcal{G},\label{eq:phi_F_phi_b}\\
        &\Y_g = \overline{\mathsf{vec}}(\mathbf{B}\mathbf{P}\bar{\mathbf{y}}_g), ~\forall g \in \mathcal{G},\label{eq:phi_F_phi_c}\\
        &[\bar{\mathbf{y}}_g]_{m'} \in \mathcal{Y},~\forall g \in \mathcal{G}, ~\forall m' \in \mathcal{M}',\label{eq:phi_F_phi_d}
    \end{align}
\end{subequations}
where $\mathcal{M'}=\{1,\ldots,U\}$. To address this, we propose a double-loop iterative algorithm, referred to as the MM-ADMM algorithm. In the outer loop, we utilize the MM method \cite{MM} to linearize objective (\ref{eq:phi_F_phi_re_a}). Then, in the inner loop, we employ the ADMM method \cite{boyd_admm} to effectively handle the complex constraints. The details of the proposed algorithm are described as follows.

\subsubsection{MM Approximation}
The main idea of the MM method for solving (\ref{eq:phi_F_phi_reformulate}) is to replace the objective function with a locally tight lower-bound surrogate function, usually a concave one, such that the resulting problem is easier to tackle. For (\ref{eq:phi_F_phi_re_a}), a simple and common approach for this approximation is to use first-order Taylor expansion of $\Tilde{\bm\phi}^H \mathbf{F}\Tilde{\bm\phi}$. Defining $g(\Tilde{\bm\phi}) = \Tilde{\bm\phi}^H \mathbf{F}\Tilde{\bm\phi}$, at iteration $t$, we have 
\begin{equation}
    g(\Tilde{\bm\phi}) \geq g(\Tilde{\bm\phi_t}) + \Re\{\nabla g(\Tilde{\bm\phi_t})^H (\Tilde{\bm\phi}-\Tilde{\bm\phi}_t) \},
\end{equation}
where $\nabla g(\Tilde{\bm\phi_t})=2\mathbf{F}\Tilde{\bm\phi}_t$. Hence, at iteration $t+1$, instead of maximizing $g(\Tilde{\bm\phi})+ 2\Re\{\boldsymbol{l}^H \Tilde{\bm\phi}\}$, we approximate it by maximizing $\Re\{(\mathbf{F}\Tilde{\bm\phi}_t+\boldsymbol{l})^H\Tilde{\bm\phi}\}$, resulting in the following problem\begin{subequations}\label{eq:f_phi}
    \begin{align}
        \max_{\bar{\bm\phi},\{\mathbf{Y}_g\},\{\bar{\mathbf{y}}_g\}}~~&\Re{\{\bar{\mathbf{f}}_t^H\bar{\bm\phi}}\}\label{eq:f_phi_a}\\
        \mathrm{s.t.}~~~~~~~&\text{(\ref{eq:phi_F_phi_b}) -- (\ref{eq:phi_F_phi_d})},\label{eq:f_phi_b}
    \end{align}
\end{subequations}
where we extract all entries of interest in $\mathbf{\Phi}$ to construct $\bar{\bm{\phi}} = [\bar{\bm{\phi}}_1^T,\ldots,\bar{\bm{\phi}}_G^T]^T$ with each $\bar{\bm{\phi}}_g \in \C^{\frac{\bar{M}(\bar{M}+1)}{2} \times 1}$ containing the diagonal and upper-triangular entries of $\mathbf{\Phi}_g$, i.e., $\bm{\phi}_g = \mathsf{vec}(\mathbf{\Phi}_g) = \mathbf{P}\bar{\bm{\phi}}_g,~\forall g\in\mathcal{G}$. The matrix $\mathbf{P}$ used to extract the diagonal and upper-triangular entries of $\mathbf{\Phi}_g$ is the same as defined in (\ref{eq:P_group}). $\bar{\mathbf{f}}_t$ is a given vector constructed based on $\mathbf{F}$, $\tilde{\bm{\phi}}_t$, $\boldsymbol{l}$, and $\mathbf{P}$ to align with the structure of $\bar{\bm{\phi}}$. Defining $\Tilde{\mathbf{f}}_t  \triangleq \mathbf{F}\Tilde{\bm\phi}_t+\boldsymbol{l}$, $\Tilde{\mathbf{F}}_t \triangleq \overline{\mathsf{vec}}(\Tilde{\mathbf{f}}_t)$, and $\mathbf{P}_G \triangleq \mathbf{I}_G \otimes \mathbf{P}$, we can prove the transformation in (\ref{eq:f_phi_a}) as follows:
\begin{align}
    (\mathbf{F}\Tilde{\bm\phi}_t+\boldsymbol{l})^H\Tilde{\bm\phi} &= \Tilde{\mathbf{f}}_t^H \mathsf{vec}(\bm\Phi)\nonumber\\
    &= \mathsf{tr}(\Tilde{\mathbf{F}}_t^H\bm\Phi)\nonumber\\
    &= \sum_{g \in \mathcal{G}} \mathsf{tr}(\Tilde{\mathbf{F}}^H_{t,g}\bm\Phi_g)\nonumber\\
    &=\sum_{g \in \mathcal{G}}\mathsf{vec}^H(\Tilde{\mathbf{F}}_{t,g}) \mathsf{vec}(\bm\Phi_g)\nonumber\\
    &= \begin{bmatrix} \mathsf{vec}^H(\Tilde{\mathbf{F}}_{t,1}),\ldots,\mathsf{vec}^H(\Tilde{\mathbf{F}}_{t,G})
    \end{bmatrix}\mathbf{P}_G\bar{\bm\phi},\label{eq:f_bar_phi_bar}
\end{align}
where each $\Tilde{\mathbf{F}}_{t,g}$ represents a submatrix of $\Tilde{\mathbf{F}}_t$, specifically denoted as $[\Tilde{\mathbf{F}}_t]_{(g-1)\bar{M}+1 : g\bar{M},(g-1)\bar{M}+1 : g\bar{M}}$. Defining $\bar{\mathbf{f}}_t$ as $\bar{\mathbf{f}}_t^H \triangleq \begin{bmatrix} \mathsf{vec}^H(\Tilde{\mathbf{F}}_{t,1}), \ldots,\mathsf{vec}^H(\Tilde{\mathbf{F}}_{t,G}) \end{bmatrix} \mathbf{P}_G$, then we have the equality $(\mathbf{F}\Tilde{\bm\phi}_t+\boldsymbol{l})^H\Tilde{\bm\phi}=\bar{\mathbf{f}}_t^H\bar{\bm\phi}$. 

\subsubsection{ADMM Methodology}\label{sec:admm_method}
Following the MM approximation in the outer loop, the ADMM method is applied in the inner loop to deal with the nonlinear coupling constraint (\ref{eq:phi_F_phi_b}). We first note that when the variable $\Y_g /\bm \Phi_g$ is fixed, constraint (\ref{eq:phi_F_phi_b}) is linear in $\bm\Phi_g /\Y_g$, as written below: 
\begin{align}
    (Y_0\mathbf{I}_{\bar{M}} + \Y_g)\bm \Phi_g &= Y_0\mathbf{I}_{\bar{M}} - \Y_g,~\forall g \in \mathcal{G},\label{eq:theta_form_admm}\\
    \Y_g(\bm \Phi_g + \mathbf{I}_{\bar{M}})&= Y_0(\mathbf{I}_{\bar{M}} - \bm \Phi_g), ~\forall g \in \mathcal{G}.\label{eq:Y_form_admm}
\end{align}
By incorporating the constraint (\ref{eq:phi_F_phi_c}) and utilizing the vectorization property, (\ref{eq:theta_form_admm}) and (\ref{eq:Y_form_admm}) can be transformed into linear equations over $\bar{\bm \phi}_g$ and $\bar{\mathbf{y}}_g$, respectively, as follows\footnote{We use $\mathbf{C}_g(\bar{\mathbf{y}}_g)$ and $\mathbf{c}_g(\bar{\mathbf{y}}_g)$ to emphasize that $\mathbf{C}_g$ and $\mathbf{c}_g$ depend on $\bar{\mathbf{y}}_g$, and use $\mathbf{D}_g(\bar{\bm\phi}_g)$ and $\mathbf{d}_g(\bar{\bm\phi}_g)$ to emphasize the dependence of $\mathbf{D}_g$ and $\mathbf{d}_g$ on $\bar{\bm\phi}_g$.}:
\begin{align}
    \mathbf{C}_g(\bar{\mathbf{y}}_g) \bar{\bm \phi}_g &= \mathbf{c}_g(\bar{\mathbf{y}}_g),~\forall g \in \mathcal{G},\label{eq:phi_form_1}\\
    \mathbf{D}_g(\bar{\bm\phi}_g) \bar{\mathbf{y}}_g &= \mathbf{d}_g(\bar{\bm\phi}_g),~\forall g \in \mathcal{G},\label{eq:y_form_1}
\end{align}where $\mathbf{C}_g(\bar{\mathbf{y}}_g) \triangleq (\mathbf{I}_{\bar{M}} \otimes (Y_0\mathbf{I}_{\bar{M}} + \overline{\mathsf{vec}}(\mathbf{B}\mathbf{P}\bar{\mathbf{y}}_g)))\mathbf{P}$, $\mathbf{c}_g(\bar{\mathbf{y}}_g) \triangleq \mathsf{vec}(Y_0\mathbf{I}_{\bar{M}}-\overline{\mathsf{vec}}(\mathbf{B}\mathbf{P}\bar{\mathbf{y}}_g))$, $\mathbf{D}_g(\bar{\bm\phi}_g) \triangleq (((\overline{\mathsf{vec}}(\mathbf{P}\bar{\bm{\phi}}_g) + \mathbf{I}_{\bar{M}})^T \otimes \mathbf{I}_{\bar{M}})\mathbf{B}\mathbf{P}$, and $\mathbf{d}_g(\bar{\bm\phi}_g) \triangleq \mathsf{vec}(Y_0\mathbf{I}_{\bar{M}}-\overline{\mathsf{vec}}(\mathbf{P}\bar{\bm{\phi}}_g))$. In the following ADMM framework, (\ref{eq:phi_form_1}) will be employed for updating the variable $\bar{\bm\phi}$, while (\ref{eq:y_form_1}) will be used for updating the variable $\bar{\mathbf{y}}_g$. However, updating $\bar{\mathbf{y}}_g$ meets a challenge due to the non-separable nature between entries of $\bar{\mathbf{y}}_g$ in (\ref{eq:y_form_1}), which makes it impossible to individually optimize each entry in $\bar{\mathbf{y}}_g$ under the non-convex constraint (\ref{eq:phi_F_phi_d}). To overcome this, one effective solution is to introduce an auxiliary variable $\mathbf{z}$ by setting $\mathbf{z} = \bar{\mathbf{y}}$, where $\bar{\mathbf{y}} = [\bar{\mathbf{y}}_1^T, \ldots, \bar{\mathbf{y}}_G^T]^T$, thereby decoupling (\ref{eq:y_form_1}) from the troublesome constraint (\ref{eq:phi_F_phi_d}). Hence, problem (\ref{eq:f_phi}) is equivalently re-written as\begin{subequations}\label{eq:f_phi_plus_z}
    \begin{align}  \max_{\bar{\bm\phi},\bar{\mathbf{y}},\mathbf{z}}~~&\Re{\{\bar{\mathbf{f}}_t^H\bar{\bm\phi}}\}\label{eq:f_phi_plus_z_a}\\
        \mathrm{s.t.}~~&\text{(\ref{eq:phi_form_1}) or (\ref{eq:y_form_1})},\label{eq:f_phi_plus_z_b}\\
        &\bar{\mathbf{y}} = \mathbf{z},\label{eq:f_phi_plus_z_c}\\
        &[\mathbf{z}]_{\Tilde{m}} \in \mathcal{Y}, ~\forall \Tilde{m} \in \Tilde{\mathcal{M}},\label{eq:f_phi_plus_z_d}
    \end{align}
\end{subequations}
where $\Tilde{\mathcal{M}}=\{1,\ldots,GU\}$. Following the ADMM framework, we construct the augmented Lagrangian (AL) function in its scaled form \cite{boyd_admm}, given by\begin{equation}
\begin{aligned}
    \mathcal{L}_{\rho}(\bar{\bm \phi},\bar{\mathbf{y}},&\mathbf{z},\bm \lambda_1,\bm \lambda_2)
    = -\Re{\{\bar{\mathbf{f}}_t^H\bar{\bm\phi}}\}
    + \frac{\rho_1}{2} \|\mathbf{C} \bar{\bm \phi} - \mathbf{c} + \bm \lambda_1\|_2^2 \\
    &+ \frac{\rho_2}{2} \|\mathbf{z} - \bar{\mathbf{y}} + \bm \lambda_2\|_2^2 - \frac{\rho_1}{2}\| \bm \lambda_1\|_2^2 - \frac{\rho_2}{2}\| \bm \lambda_2\|_2^2,\label{eq:MM_AL_sumrate}
\end{aligned}   
\end{equation}
where $\mathbf{C} = \mathsf{blkdiag}(\mathbf{C}_1,\ldots,\mathbf{C}_G)$, $\mathbf{c}=[\mathbf{c}_1^T,\ldots,\mathbf{c}_G^T]^T$, and $\bm \lambda_1 \in \C^{G\bar{M}^2 \times 1}$, $\bm \lambda_2 \in \C^{U \times 1}$ denote the scaled Lagrange multipliers corresponding to constraints (\ref{eq:f_phi_plus_z_b}) and (\ref{eq:f_phi_plus_z_c}), respectively, and $\rho_1$ and $\rho_2$ are penalty coefficients.

The ADMM algorithm alternately updates the primal variables $\bar{\bm \phi}$, $\bar{\mathbf{y}}$ and $\mathbf{z}$, as well as the dual variables $\bm \lambda_1$ and $\bm \lambda_2$, while keeping $\rho_1$ and $\rho_2$ fixed. This process iterates until convergence. The detailed algorithm is given below.
\begin{itemize}
    \item Update $\bar{\bm \phi}$ by solving
    \begin{equation}\label{eq:sub_phi_bar_mm}
    \begin{aligned}
        \min_{\bar{\bm \phi}}~-\Re{\{\bar{\mathbf{f}}_t^H\bar{\bm\phi}}\}
         + \frac{\rho_1}{2} \|\mathbf{C} \bar{\bm \phi} - \mathbf{c} + \bm \lambda_1\|_2^2,
    \end{aligned}
    \end{equation}
    where (\ref{eq:sub_phi_bar_mm}) is an unconstrained convex optimization and the optimal solution is given by
    \begin{equation}\label{eq:sub_phi_bar_opt_mm}
    \bar{\bm \phi} = \left(\rho_1 \mathbf{C}^H\mathbf{C}\right)^{-1} \left(\bar{\mathbf{f}}_t + \rho_1\mathbf{C}^H(\mathbf{c} - \bm \lambda_1)\right);
    \end{equation}
    \item Update $\bar{\mathbf{y}}$ by solving
    \begin{equation}\label{eq:sub_y_bar_mm}
        \min_{\bar{\mathbf{y}}}~\frac{\rho_1}{2} \|\mathbf{D}\bar{\mathbf{y}}-\mathbf{d}+\bm \lambda_1\|_2^2
        +\frac{\rho_2}{2} \|\mathbf{z} - \bar{\mathbf{y}} + \bm \lambda_2\|_2^2,
    \end{equation}
    where $\mathbf{D} = \mathsf{blkdiag}(\mathbf{D}_1,\ldots,\mathbf{D}_G)$ and $\mathbf{d}=[\mathbf{d}_1^T,\ldots,\mathbf{d}_G^T]^T$. Problem (\ref{eq:sub_y_bar_mm}) is an unconstrained convex optimization and the optimal solution is given by
    \begin{equation}\label{eq:sub_y_bar_opt_mm}
    \begin{aligned}
        \bar{\mathbf{y}} = \Big(\rho_1\mathbf{D}^H\mathbf{D}&+\rho_2\mathbf{I}_{GU}\Big)^{-1}\\
        &\times\left(\rho_1\mathbf{D}^H(\mathbf{d}-\bm \lambda_1)+\rho_2(\mathbf{z}+\bm \lambda_2)\right);
    \end{aligned}
    \end{equation}
    \item Update $\mathbf{z}$ by solving
    \begin{equation}\label{eq:sub_z_mm}
        \min_{\mathbf{z}}~\|\mathbf{z} - \bar{\mathbf{y}} + \bm \lambda_2\|_2^2 ~~
        \mathrm{s.t.} ~~\text{(\ref{eq:f_phi_plus_z_d})}.
    \end{equation}
    Since constraint (\ref{eq:f_phi_plus_z_d}) is separable with respect to each entry of vector $\mathbf{z}$, problem (\ref{eq:sub_z_mm}) can be decomposed into $GU$ independent one-dimensional problems. Defining $\mathbf{r} \triangleq \bar{\mathbf{y}}-\bm \lambda_2$, each one-dimensional problem can be expressed as
    \begin{subequations}\label{eq:sub_z_1D}
    \begin{align}
        \min_{[\mathbf{z}]_{\Tilde{m}}}~~&\left|[\mathbf{z}]_{\Tilde{m}} - [\mathbf{r}]_{\Tilde{m}}\right|^2\label{eq:sub_z_1D_a}\\
        \mathrm{s.t.}~~&[\mathbf{z}]_{\Tilde{m}} \in \mathcal{Y}.\label{eq:sub_z_1D_b}
    \end{align}
    \end{subequations}
    Substituting (\ref{eq:sub_z_1D_b}) into the objective function (\ref{eq:sub_z_1D_a}) further yields 
    \begin{equation}\label{eq:sub_z_1D_reformulate}
        \min_{\theta_{\Tilde{m}}}~\left|\beta e^{\jmath \theta_{\Tilde{m}}} - |\psi|e^{\jmath \arg \psi}\right|^2 ~~
        \mathrm{s.t.} ~~\theta_{\Tilde{m}} \in [\theta_\mathrm{min}, \theta_\mathrm{max}],
    \end{equation}
    where $\psi\triangleq[\mathbf{r}]_{\Tilde{m}}-\alpha$. To minimize the objective function in (\ref{eq:sub_z_1D_reformulate}), the angle $\theta_{\Tilde{m}}$ should be as close as possible to $\arg \psi$. Thus the optimal solution is given by 
    \begin{equation}\label{eq:theta_each_solution}
     \begin{aligned}
        \theta_{\Tilde{m}} = 
        \begin{cases}
            \theta_\mathrm{min}, & \text{if}~\arg \psi \leq \theta_\mathrm{min};\\
            \theta_\mathrm{max}, & \text{if}~\arg \psi \geq \theta_\mathrm{max};\\
            \arg \psi, & \text{if}~\theta_\mathrm{min} < \arg \psi < \theta_\mathrm{max};
        \end{cases}
    \end{aligned}
    \end{equation}
    \item Update $\bm \lambda_1$ and $\bm \lambda_2$ as
    \begin{align}
        \bm \lambda_1 &= \bm \lambda_1 + \mathbf{C}\bar{\bm \phi} - \mathbf{c},\label{eq:update_lambda1}\\
        \bm \lambda_2 &= \bm \lambda_2 + \mathbf{z} -\bar{\mathbf{y}}.\label{eq:update_lambda2}
    \end{align}

\end{itemize}

\begin{algorithm}[t]
    \caption{Proposed MM-ADMM Algorithm}
    \label{alg:mm_admm}
    \begin{algorithmic}[1]
        \REQUIRE $h_{RT}$, $\mathbf{h}_{RI}$, $\mathbf{h}_{IT}$, $\alpha$, $\beta$, $\bar{M}$, $\theta_\mathrm{min}$, $\theta_\mathrm{max}$.
        \ENSURE $\bar{\bm \phi}^\star$.
            \STATE {Calculate $\mathbf{B}$, $\mathbf{P}$, $\mathbf{F}$, $\boldsymbol{l}$}.
            \STATE {Initialize $\bar{\mathbf{y}}$, $\Tilde{\bm \phi}_0$, $\mathbf{z}$, $\bm \lambda_1,\bm \lambda_2$ and $t = 0$.}
            \STATE {Calculate $\bar{\mathbf{f}}_{0}$ by (\ref{eq:f_bar_phi_bar}).}
            \WHILE {no convergence of objective (\ref{eq:phi_F_phi_re_a})}
                \WHILE {no convergence of objective (\ref{eq:f_phi_a})}
                    \STATE {Update $\mathbf{C}$ and $\mathbf{c}$ from $\bar{\mathbf{y}}$, and then update $\bar{\bm \phi}$ by (\ref{eq:sub_phi_bar_opt_mm}).}
                    \STATE {Update $\mathbf{D}$ and $\mathbf{d}$ from $\bar{\bm \phi}$, and then update $\bar{\mathbf{y}}$ by (\ref{eq:sub_y_bar_opt_mm}).}   
                    \STATE {Update $[\mathbf{z}]_{\Tilde{m}},~\forall \Tilde{m} \in \Tilde{\mathcal{M}}$  by (\ref{eq:theta_each_solution}).}
                    \STATE {Update $\bm \lambda_1$ and $\bm \lambda_2$ by (\ref{eq:update_lambda1}) and (\ref{eq:update_lambda2}), respectively.}
                \ENDWHILE
                \STATE {Update $\Tilde{\bm \phi}_{t}$ as $\Tilde{\bm \phi}_{t} = \mathsf{vec}(\bm \Phi)$, $\bar{\mathbf{f}}_t$ by (\ref{eq:f_bar_phi_bar}), and $t = t+1$}.
            \ENDWHILE
            \STATE {Return $\bar{\bm \phi}$ as $\bar{\bm \phi}^\star$.}
    \end{algorithmic}
\end{algorithm}

\subsubsection{Summary}The proposed MM-ADMM algorithm is summarized in Algorithm \ref{alg:mm_admm}. Once $\bar{\bm \phi}^\star$ is obtained, we can compute $\bm \Phi^\star_g$ as $\bm \Phi^\star_g = \overline{\mathsf{vec}}(\mathbf{P}\bar{\bm \phi}^\star_g), ~\forall g \in \mathcal{G}$. The final solution for $\bm \Phi^\star$ is then given by $\bm \Phi^\star = 
\mathsf{blkdiag}(\bm \Phi^\star_1,\ldots,\bm \Phi^\star_G)$.

\subsection{Low-Complexity Algorithm}\label{sec:low_complex_admm}
The proposed MM-ADMM algorithm requires a double-loop iteration, which might require high computational complexity to guarantee convergence. To ease the requirement of the double-loop iteration, we first recall that, for a lossless group-connected reconfigurable admittance network, the received signal power $\left|h_{RT} + \mathbf{h}_{RI}^H\bm \Phi \mathbf{h}_{IT} \right|^2$ reaches its upper bound when $\bm \Phi$ satisfies \cite{shen2021modelling}
\begin{equation}
    e^{\jmath\arg(h_{RT})}\bar{\mathbf{h}}_{RI,g} = \bm \Phi_g \bar{\mathbf{h}}_{IT,g}, ~\forall g \in \mathcal{G}, \label{eq:upperbound_theta}
\end{equation}
where $\bar{\mathbf{h}}_{RI}=[\bar{\mathbf{h}}_{RI,1}^T,\ldots,\bar{\mathbf{h}}_{RI,G}^T]^T$ with $\bar{\mathbf{h}}_{RI,g} \triangleq \frac{\mathbf{h}_{RI,g}}{\| \mathbf{h}_{RI,g} \|_2} \in \C^{\bar{M} \times 1}$, and $\bar{\mathbf{h}}_{IT}=[\bar{\mathbf{h}}_{IT,1}^T,\ldots,\bar{\mathbf{h}}_{IT,G}^T]^T$ with $\bar{\mathbf{h}}_{IT,g} \triangleq \frac{\mathbf{h}_{IT,g}}{\| \mathbf{h}_{IT,g} \|_2} \in \C^{\bar{M} \times 1}$. By substituting $\bm \Phi_g$ with $\Y_g,~\forall g \in \mathcal{G}$ based on (\ref{eq:phi_F_phi_b}), (\ref{eq:upperbound_theta}) can be equivalently re-written as
\begin{equation}
    \Y \left(\breve{\mathbf{h}}_{RI} + \bar{\mathbf{h}}_{IT} \right) - Y_0 \left( \bar{\mathbf{h}}_{IT} - \breve{\mathbf{h}}_{RI}\right) = \mathbf{0}.\label{eq:opt_Y_simplify}
\end{equation}
where $\breve{\mathbf{h}}_{RI}\triangleq e^{\jmath\arg(h_{RT})}\bar{\mathbf{h}}_{RI}$. Note that the above conditions hold true for lossless BD-RISs with unitary scattering matrix and purely imaginary admittance matrix, while they do not hold true for lossy BD-RISs due to the additional constraint on the admittance matrix $\mathbf{Y}$. Therefore, we propose minimizing the 2-norm of the left-hand side of (\ref{eq:opt_Y_simplify}) as an approximation to maximize the received signal power. This leads to the following constrained linear least squares problem:
\begin{subequations}\label{eq:Ya-b}
    \begin{align}
        \min_{\{\mathbf{Y}_g\},\{\bar{\mathbf{y}}_g\}}~& \|\Y \mathbf{a} - \mathbf{b}\|_2^2 \label{eq:Ya-b_a}\\
        \mathrm{s.t.}~~~~
        &\text{(\ref{eq:blkdiag}), (\ref{eq:phi_F_phi_c}), (\ref{eq:phi_F_phi_d})},\label{eq:Ya-b_b}
    \end{align}
\end{subequations}
where $\mathbf{a} \triangleq \breve{\mathbf{h}}_{RI} + \bar{\mathbf{h}}_{IT}$ and $\mathbf{b} \triangleq Y_0 \left( \bar{\mathbf{h}}_{IT} - \breve{\mathbf{h}}_{RI}\right)$. Problem (\ref{eq:Ya-b}) can be equivalently reformulated by eliminating constraints (\ref{eq:blkdiag}) and (\ref{eq:phi_F_phi_c}), resulting in the following form:
\begin{subequations}\label{eq:Ay-b}
    \begin{align}
        \min_{\bar{\mathbf{y}}}~~&\|\mathbf{A}\bar{\mathbf{y}} - \mathbf{b}\|_2^2 \label{eq:Ay-b_a}\\
        \mathrm{s.t.}~~
        &[\bar{\mathbf{y}}]_{\Tilde{m}} \in \mathcal{Y}, ~\forall \Tilde{m} \in \Tilde{\mathcal{M}},\label{eq:Ay-b_b}
    \end{align}
\end{subequations}
where $\mathbf{A} \in \C^{M \times GU}$ is a block-diagonal matrix given by $\mathbf{A} = \mathsf{blkdiag}(\mathbf{A}_1,\ldots,\mathbf{A}_G)$, with $\mathbf{A}_g \in \C^{\bar{M} \times U}, ~\forall g \in \mathcal{G}$ derived from vector $\mathbf{a}_g$ through a series of transformations. Each $\mathbf{a}_g \in \C^{\bar{M} \times 1}$ represents a block of the vector $\mathbf{a}$, which is defined as $\mathbf{a} = [\bar{\mathbf{a}}^T_1, \ldots, \bar{\mathbf{a}}^T_G]^T$. To prove the transformation in (\ref{eq:Ay-b_a}), we utilize the vectorization property to reformulate $\Y_g \mathbf{a}_g, ~\forall g \in \mathcal{G}$ as
\begin{equation}
    \Y_g\mathbf{a}_g = \Tilde{\mathbf{A}}_g\mathsf{vec}(\Y_g), \label{eq:Yg_ag_vec}
\end{equation}
where $\Tilde{\mathbf{A}}_g \triangleq \mathbf{a}_g^T \otimes \mathbf{I}_{\bar{M}}$. Re-write $\mathsf{vec}(\Y_g)$ as $\mathsf{vec}(\Y_g)=\mathbf{B} \mathbf{P}\bar{\mathbf{y}}_g$, equation (\ref{eq:Yg_ag_vec}) therefore simplifies to
\begin{equation}
    \Y_g\mathbf{a}_g = \mathbf{A}_g\bar{\mathbf{y}}_g, ~\forall g \in \mathcal{G},\label{eq:Yg_ag_vec_simplify}
\end{equation}
where $\mathbf{A}_g \triangleq \Tilde{\mathbf{A}}_g \mathbf{B} \mathbf{P}$. Given the block-diagonal structure of both $\Y$ and $\mathbf{A}$, along with the column-wise stacking of $\mathbf{a}$ and $\bar{\mathbf{y}}$, we establish that $\Y \mathbf{a} = \mathbf{A} \bar{\mathbf{y}}$.

The main challenge in solving problem (\ref{eq:Ay-b}) still arises from the non-separability of the objective function (\ref{eq:Ay-b_a}) with respect to $\bar{\mathbf{y}}$ under constraint (\ref{eq:Ay-b_b}). We again introduce a copy $\mathbf{z}$ of $\bar{\mathbf{y}}$ to decouple (\ref{eq:Ay-b_a}) from (\ref{eq:Ay-b_b}). Therefore, problem (\ref{eq:Ay-b}) can be equivalently re-written as
\begin{subequations}\label{eq:Ay-b_z}
    \begin{align}
        \min_{\bar{\mathbf{y}},\mathbf{z}}~~& \|\mathbf{A}\bar{\mathbf{y}} - \mathbf{b}\|_2^2 \label{eq:Ay-b_z_a}\\
        \mathrm{s.t.}~~
        &\text{(\ref{eq:f_phi_plus_z_c}), (\ref{eq:f_phi_plus_z_d})}.\label{eq:Ay-b_z_b}
    \end{align}
\end{subequations}
We apply the ADMM method \cite{boyd_admm} to solve (\ref{eq:Ay-b_z}). The AL function of (\ref{eq:Ay-b_z}) is given by:
\begin{equation}
    \mathcal{L}_{\rho}(\bar{\mathbf{y}},\mathbf{z},\mathbf{u}) = \|\mathbf{A}\bar{\mathbf{y}} - \mathbf{b}\|_2^2 + \frac{\rho}{2} \|\mathbf{z} - \bar{\mathbf{y}} + \mathbf{u}\|_2^2 - \frac{\rho}{2} \|\mathbf{u}\|_2^2, \label{eq:least_AL}
\end{equation}
where $\mathbf{u} \in \C^{GU \times 1}$ is the scaled Lagrangian multiplier and $\rho$ is the penalty coefficient. The detailed algorithm is presented below.
\begin{itemize}
    \item Update $\bar{\mathbf{y}}$ by solving \begin{equation}\label{eq:admm_siso_y_bar_1}
    \min_{\bar{\mathbf{y}}}~\|\mathbf{A}\bar{\mathbf{y}} - \mathbf{b}\|_2^2 + \frac{\rho}{2} \|\mathbf{z} - \bar{\mathbf{y}} + \mathbf{u}\|_2^2,
    \end{equation}
    where (\ref{eq:admm_siso_y_bar_1}) is an unconstrained convex optimization and the optimal solution is 
    \begin{equation}\label{eq:admm_siso_y_bar_2}
    \bar{\mathbf{y}} = \left(2\mathbf{A}^H\mathbf{A} + \rho\mathbf{I}_{GU}\right)^{-1} \left(2\mathbf{A}^H\mathbf{b} + \rho(\mathbf{z}+\mathbf{u})\right);
    \end{equation}
    \item Update $\mathbf{z}$ by solving
    \begin{equation}\label{eq:admm_siso_z}
        \min_{\mathbf{z}}~\|\mathbf{z} - \bar{\mathbf{y}} + \mathbf{u}\|_2^2
        ~~\mathrm{s.t.}~~\text{(\ref{eq:f_phi_plus_z_d})}.
    \end{equation}
    The structure of problem (\ref{eq:admm_siso_z}) is the same as (\ref{eq:sub_z_mm}) and its solution is provided in (\ref{eq:theta_each_solution}), with $\bm \lambda_2$ replaced by $\mathbf{u}$;
    \item Update $\mathbf{u}$ as
    \begin{equation}
        \mathbf{u}=\mathbf{u} + \mathbf{z} -\bar{\mathbf{y}}.\label{eq:update_u}
    \end{equation}

\end{itemize}

With the solution $\bar{\mathbf{y}}^\star$ to problem (\ref{eq:Ay-b_z}), we can compute  $\Y_g^\star, ~\forall g \in \mathcal{G}$ as $\Y_g^\star = \overline{\mathsf{vec}}(\mathbf{B}\mathbf{P}\bar{\mathbf{y}}_g^\star)$. The admittance matrix is then given by $\mathbf{Y}^\star = \mathsf{blkdiag}(\mathbf{Y}_{1}^\star,\ldots,\mathbf{Y}_{G}^\star)$, from which the scattering matrix $\bm\Phi^\star$ is obtained using (\ref{eq:Phi_Y}).

\textit{Remark 3:} The only difference between group-connected and forest-connected architectures lies in the number of variables in each $\bar{\Y}_g$. For the group-connected BD-RIS, $\bar{\Y}_g,~\forall g\in\mathcal{G}$ is a matrix with no zero entries, whereas certain entries in $\bar{\Y}_g$ are zero according to (\ref{eq:Y_tri}) or (\ref{eq:Y_arr}) for forest-connected BD-RIS. This difference affects our optimization algorithms only through the mapping $\bar{\Y}_g=\overline{\mathsf{vec}}(\mathbf{P}\bar{\mathbf{y}}_g)$, where matrix $\mathbf{P}$ differs for the group-connected and forest-connected BD-RISs, as specified in (\ref{eq:P_group}) – (\ref{eq:P_arr}); each $\bar{\mathbf{y}}_g$ contains $\frac{\bar{M}(\bar{M}+1)}{2}$ entries for the group-connected BD-RIS and $2\bar{M}-1$ entries for the forest-connected BD-RIS. Apart from this modification, both the MM-ADMM and the low-complexity algorithm apply in the same way to the two architectures.

\subsection{Complexity Analysis}
In this subsection, we analyze the computational complexity of the proposed algorithms. The MM-ADMM algorithm (Algorithm \ref{alg:mm_admm}) employs a double-loop iterative design. In the inner loop, each iteration updates $\bar{\bm \phi}$, $\bar{\mathbf{y}}$ and $\mathbf{z}$, requiring a total of $\mathcal{O}\{\bar{M}^3 M^3\}$ operations. In the outer loop, the update of $\bar{\mathbf{f}}_t$ per iteration demands $\mathcal{O}\{M^4\}$ operations. Therefore, the overall complexity of Algorithm \ref{alg:mm_admm} is $\mathcal{O}\{I_\mathrm{out}(M^4 + I_\mathrm{in}\bar{M}^3 M^3)\}$, where $I_\mathrm{out}$ and $I_\mathrm{in}$ represent the number of iterations in the outer and inner loops, respectively. For comparison, the low-complexity algorithm employs a single-loop ADMM framework, resulting in an overall complexity of $\mathcal{O}\{I_\mathrm{sig}\bar{M}^3 M^3\}$, where $I_\mathrm{sig}$ denotes the number of iterations.

\textit{Remark 4:} The overall complexity of the MM-ADMM algorithm is identical for both group- and forest-connected BD-RISs. This is because, in forest-connected BD-RIS, each iteration requires $\mathcal{O}\{\bar{M}^3 M^3\}$ operations to update $\bar{\bm \phi}$. For forest-connected BD-RIS, the low-complexity algorithm further reduces overall complexity: updating $\bar{\mathbf{y}}$ now requires $\mathcal{O}\{M^3\}$ operations instead of $\mathcal{O}\{\bar{M}^3 M^3\}$, leading to a total complexity of $\mathcal{O}\{I_\mathrm{sig}M^3\}$.

\textit{Remark 5:} The main advantage of the low-complexity algorithm lies in its substantially reduced computational complexity compared to the MM-ADMM algorithm. Due to the non-convexity of the objective function, the exact MM-ADMM solution is obtained in a double-loop structure, where the outer loop is to deal with the non-convex objective, while the inner loop is to deal with complex constraints of lossy BD-RIS using the ADMM framework. In contrast, the low-complexity algorithm directly constructs a convex objective function and thus needs only a single loop ADMM. This reduces the complexity from $\mathcal{O}\{I_\mathrm{out}(M^4 + I_\mathrm{in}\bar{M}^3 M^3)\}$ to $\mathcal{O}\{I_\mathrm{sig}\bar{M}^3 M^3\}$ for group-connected BD-RIS, and further to $\mathcal{O}\{I_\mathrm{sig}M^3\}$ for forest-connected BD-RIS. However, since it solves the approximate problem (\ref{eq:Ay-b}) rather than the original power maximization problem, performance degradation may occur when the approximation is inaccurate, e.g., for D-RIS. This complexity–performance trade-off will be quantified in Section \ref{sec:simulation}.

\section{Optimization for BD-RIS-Aided MU-MISO System}\label{sec:mu_miso}
In this section, we first present the system model with the lossy BD-RIS modeling in MU-MISO systems. Then we formulate the sum-rate maximization problem and propose an iterative algorithm to jointly optimize the transmit precoder and the BD-RIS matrix.

\subsection{System Model and Problem Formulation}
We then consider an $M$-element BD-RIS-aided MU-MISO system with an $N$-antenna base station (BS) and $K$ single-antenna users. At the BS, it is assumed that exact and instantaneous channel state information (CSI) is available. Let $\mathbf{s} \triangleq [s_1, \ldots, s_K]^T \in \C^{K \times 1}$ represent the transmit symbol vector with $\mathbb{E}\{\mathbf{s}\mathbf{s}^H\} = \mathbf{I}_K$, and $\mathbf{W} \triangleq [\mathbf{w}_1, \ldots, \mathbf{w}_K] \in \C^{N \times K}$ denote the precoder matrix at the BS. Here, $s_k \in \C$ and $\mathbf{w}_k \in \C^{N \times 1}$ are the transmit symbol and precoding vector for user $k$, respectively, where $k \in \mathcal{K} = \{1, \ldots, K\}$. At each user, the received signal is given by
\begin{equation}\label{eq:y_k}
\begin{aligned}
    y_k = &~(\mathbf{h}_{RT,k}^H + \mathbf{h}_{RI,k}^H\bm \Phi\mathbf{H}_{IT})\mathbf{w}_k s_k \\
    &+ (\mathbf{h}_{RT,k}^H + \mathbf{h}_{RI,k}^H\bm \Phi\mathbf{H}_{IT}) \sum_{p \in \mathcal{K}, p \neq k}\mathbf{w}_p s_p + n_k, ~\forall k \in \mathcal{K},
\end{aligned} 
\end{equation}
where $\mathbf{h}_{RT,k} \in \C^{N \times 1}$, $\mathbf{h}_{RI,k} \in \C^{M \times 1}$, and $\mathbf{H}_{IT} \in \C^{M \times N}$ denote the channel matrices from BS to user $k$, from BD-RIS to user $k$, and from BS to BD-RIS, respectively, and $n_{k} \sim \mathcal{CN}(0, \sigma_{k}^2)$ denotes the AWGN for user $k$, where $k \in \mathcal{K}$. Define $\Tilde{\mathbf{h}}_k \triangleq (\mathbf{h}_{RT,k}^H + \mathbf{h}_{RI,k}^H\bm \Phi\mathbf{H}_{IT})^H, ~\forall k \in \mathcal{K}$. The signal-to-interference-plus-noise ratio (SINR) for each user can be written as
\begin{equation}\label{eq:sinr}
    \gamma_k = \frac{\left|\Tilde{\mathbf{h}}_k^H \mathbf{w}_k \right|^2}{\sum_{p \in \mathcal{K}, p \neq k} \left|\Tilde{\mathbf{h}}_k^H \mathbf{w}_p \right|^2 + \sigma_k^2}, ~\forall k \in \mathcal{K}.
\end{equation}

We aim to jointly design the transmit precoder and the BD-RIS matrix to maximize the sum-rate of the MU-MISO system, taking into account the losses in BD-RIS and the transmit power constraint at the BS. Hence, the problem is formulated as\footnote{The symmetric constraint of $\bm \Phi$ in (\ref{eq:p1_b}) ensures that the symmetry of $\bm \Phi$ is maintained during the alternating updates of $\bar{\bm\phi}$ and $\bar{\mathbf{y}}$ in Section \ref{sec:update_Phi}.}
\begin{subequations}
    \label{eq:problem1}
    \begin{align}
        \label{eq:obj1}
        \max_{\mathbf{W}, \bm \Phi} ~~ &f_1(\mathbf{W}, \bm \Phi) = \sum_{k\in \mathcal{K}}\log_2(1 + \gamma_{k})\\
        \mathrm{s.t.} ~~ &\label{eq:p1_b}
        \bm \Phi = \bm \Phi^T,\\
        \label{eq:p1_c}
        & \text{(\ref{eq:phi_F_phi_b}) -- (\ref{eq:phi_F_phi_d})},\\
        \label{eq:p1_d}
        &\|\mathbf{W}\|_F^2 \le P,
    \end{align}
\end{subequations}
where $P$ is the total transmit power at the BS. Problem (\ref{eq:problem1}) involves the same lossy BD-RIS constraints as problem (\ref{eq:received_signal_power_max}) in the SISO case. In contrast to the SISO systems, where only the BD-RIS matrix is optimized, we should jointly design the transmit precoder and BD-RIS matrix in MU-MISO systems. Furthermore, the ratio terms within the logarithm of the objective function in problem (\ref{eq:problem1}) complicates the optimization. To tackle these challenges, in the following subsections, we first reformulate the objective function (\ref{eq:obj1}) using FP theory \cite{FPI,FPII}, converting (\ref{eq:problem1}) into a multi-block optimization problem, and then optimize all variables iteratively.

\subsection{FP Reformulation}
To deal with the complicated sum-rate objective function, we employ the FP technique proposed in \cite{FPI,FPII}, which, by introducing auxiliary variables $\nu_{k}$ and $\tau_k,~\forall k \in \mathcal{K}$, transforms the original problem (\ref{eq:problem1}) into the following form: 
\begin{subequations}
    \label{eq:fp_f_final}
    \begin{align}
        \label{eq:obj_f2}
        \max_{\mathbf{W},\bm \Phi, \bm \nu, \bm \tau} ~~ &f_2(\mathbf{W},\bm \Phi, \bm \nu, \bm \tau) = \sum_{k\in\mathcal{K}} \bigg(\log(1 + \nu_{k}) - \nu_{k} \nonumber\\
        &+ 
        2\sqrt{1 + \nu_{k}}\Re\{\tau_{k}^*\tilde{\mathbf{h}}_{k}^H\mathbf{w}_{k}\}\nonumber\\ 
       &- |\tau_{k}|^2 \Big(\sum_{p\in\mathcal{K}} |\tilde{\mathbf{h}}_{k}^H\mathbf{w}_{p}|^2 + \sigma_{k}^2\Big)\bigg),\\
        \label{eq:f2_b}
        \mathrm{s.t.} ~~~~~ & \text{(\ref{eq:p1_b}) -- (\ref{eq:p1_d})},
    \end{align}
\end{subequations}
where $\bm \nu \triangleq [\nu_{1}, \ldots, \nu_{K}]^T \in \mathbb{R}^{K\times 1}$ and $\bm \tau \triangleq [\tau_{1}, \ldots, \tau_{K}]^T \in \mathbb{C}^{K\times 1}$. Compared to the original problem (\ref{eq:problem1}), the above formulation is more tractable, as the objective function with respect to each variable is simple and concave. To utilize this property, we employ a BCD framework\footnote{The main difference between our BCD algorithm and the one in \cite{weighted_sum_rate_ris_Guo} lies in the update of $\bm \Phi$. The considered lossy BD-RIS must satisfy the complex constraints in (\ref{eq:phi_F_phi_b}) – (\ref{eq:phi_F_phi_d}), rather than the simple unit-modulus constraint of D-RIS in \cite{weighted_sum_rate_ris_Guo}. Hence, the algorithm in \cite{weighted_sum_rate_ris_Guo} is not applicable here.} to iteratively update each variable block until convergence. In the following subsections, we will show the details for updating each block in (\ref{eq:fp_f_final}).

\subsection{Update of $\bm \nu$, $\bm \tau$ and $\mathbf{W}$}
Given $\bm \Phi$ and $\mathbf{W}$, the updates of variables $\bm \nu$ and $\bm \tau$ follow a standard procedure as given in \cite{FPII}:
\begin{align}
    \label{eq:nu_opt}
    \nu_{k} &= \gamma_{k}, ~\forall k \in \mathcal{K},\\
    \label{eq:tau_opt}
        \tau_{k} &= \frac{\sqrt{1 + \nu_{k}}\tilde{\mathbf{h}}_{k}^H\mathbf{w}_{k}}{\sum_{p \in \mathcal{K}}|\tilde{\mathbf{h}}_{k}^H\mathbf{w}_{p}|^2 + \sigma_{k}^2}, ~\forall k \in \mathcal{K}.
\end{align}
When $\bm \Phi$, $\bm \nu$ and $\bm \tau$ are fixed, problem (\ref{eq:fp_f_final}) becomes a convex optimization problem. The solution for $\mathbf{W}$ can then be obtained using the method of Lagrange multipliers as
\begin{equation}
\label{eq:w_star}
    \mathbf{w}_{k} = \left(\sum_{p\in\mathcal{K}} |\tau_{p}|^2\tilde{\mathbf{h}}_{p}\tilde{\mathbf{h}}_{p}^H + \lambda\mathbf{I}_N\right)^{-1}\sqrt{1 + \nu_{k}}\tau_k\tilde{\mathbf{h}}_{k}, ~\forall k \in \mathcal{K},
\end{equation}
where $\lambda \ge 0$ is the Lagrange multiplier associated with the power constraint (\ref{eq:p1_d}) and can be determined via a bisection search; see detailed discussions in \cite{hybrid_hongyu,FPI}.

\subsection{Update of $\bm \Phi$}\label{sec:update_Phi}
When $\mathbf{W}$, $\bm \nu$ and $\bm \tau$ are fixed, problem (\ref{eq:fp_f_final}) with respect to $\bm \Phi$ can be written as
\begin{subequations}\label{eq:sub_theta}
    \begin{align}\label{eq:obj_theta_1}
        \min_{\bm \Phi} ~~ &\sum_{k\in\mathcal{K}}\Big(
        |\tau_{k}|^2\sum_{p\in\mathcal{K}} |\mathbf{h}_{RI,k}^H\bm \Phi\mathbf{v}_p|^2\notag\\
        &- 2\Re\{\sqrt{1 + \nu_{k}} \tau_k^*\mathbf{h}_{RI,k}^H \bm \Phi \mathbf{v}_k
        - |\tau_{k}|^2\mathbf{h}_{RI,k}^H \bm \Phi \mathbf{x}_k\}  \Big)\\
        \mathrm{s.t.} ~~ &\text{(\ref{eq:p1_b}), (\ref{eq:p1_c})},
    \end{align}
\end{subequations}
where $\mathbf{v}_p \triangleq \mathbf{H}_{IT}\mathbf{w}_p$ and $\mathbf{x}_k\triangleq\mathbf{H}_{IT}\sum_{p\in\mathcal{K}}\mathbf{w}_p\mathbf{w}_p^H\mathbf{h}_{RT,k}$, $~\forall p,k \in \mathcal{K}$. Let $\mathbf{v}_p=[\mathbf{v}_{p,1}^T,\ldots,\mathbf{v}_{p,G}^T]^T$ and $\mathbf{h}_{RI,k}^H = [\mathbf{h}_{RI,k,1}^H,\ldots,\mathbf{h}_{RI,k,G}^H]$, we have $\mathbf{h}_{RI,k}^H\bm \Phi\mathbf{v}_p = \sum_{g \in \mathcal{G}} \mathbf{h}_{RI,k,g}^H \bm \Phi_g \mathbf{v}_{p,g}$. Let $\mathbf{x}_k=[\mathbf{x}_{k,1}^T,\ldots,\mathbf{x}_{k,G}^T]^T$, and we have $\mathbf{h}_{RI,k}^H\bm \Phi\mathbf{x}_k = \sum_{g \in \mathcal{G}} \mathbf{h}_{RI,k,g}^H \bm \Phi_g \mathbf{x}_{k,g}$. We then apply the vectorization property to transform $\mathbf{h}_{RI,k,g}^H\bm \Phi_g\mathbf{v}_{p,g}$ and $\mathbf{h}_{RI,k,g}^H\bm \Phi_g\mathbf{x}_{k,g}, ~\forall g \in \mathcal{G}$, as
\begin{align}
     \mathbf{h}_{RI,k,g}^H \bm \Phi_g \mathbf{v}_{p,g} &= \mathbf{e}_{p,k,g}^H\bm \phi_g,\\
     \mathbf{h}_{RI,k,g}^H \bm \Phi_g \mathbf{x}_{k,g} &= \mathbf{m}_{k,g}^H\bm \phi_g.
\end{align}
where $\mathbf{e}_{p,k}^H = [\mathbf{e}_{p,k,1}^H,\ldots,\mathbf{e}_{p,k,G}^H] \in \C^{1 \times G\bar{M}^2}$ with $\mathbf{e}_{p,k,g}^H \triangleq \mathbf{v}_{p,g}^T \otimes \mathbf{h}_{RI,k,g}^H$, and $\mathbf{m}_{k}^H = [\mathbf{m}_{k,1}^H,\ldots,\mathbf{m}_{k,G}^H] \in \C^{1 \times G\bar{M}^2}$ with $\mathbf{m}_{k,g}^H \triangleq \mathbf{x}_{k,g}^T \otimes \mathbf{h}_{RI,k,g}^H$. Consequently, the objective function (\ref{eq:obj_theta_1}) can be reformulated as
\begin{equation}
\begin{aligned}
     \sum_{k\in\mathcal{K}}\Big(
        |\tau_{k}|^2 &\sum_{p\in\mathcal{K}} |\mathbf{e}_{p,k}^H \bm \phi|^2\\
        &- 2\Re\{(\sqrt{1 + \nu_{k}} \tau_k^*\mathbf{e}_{k,k}^H
        - |\tau_{k}|^2\mathbf{m}_k^H) \bm \phi\}  \Big),\label{eq:sub_theta_reformulate}
\end{aligned}
\end{equation}
where $\bm \phi = [\bm \phi_1^T,\ldots,\bm \phi_G^T]^T \in \C^{G\bar{M}^2 \times 1}$. The symmetric constraint (\ref{eq:p1_b}) can be eliminated by re-writing $\bm \phi$ as $\bm \phi = \mathbf{P}_G \bar{\bm \phi}$. By further simplifying (\ref{eq:sub_theta_reformulate}), problem is reformulated into the following form:
\begin{equation}\label{eq:phi_q_phi}
        \min_{\bar{\bm\phi},\{\mathbf{Y}_g\},\{\bar{\mathbf{y}}_g\}} ~  \bar{\bm \phi}^H \mathbf{Q} \bar{\bm \phi} - 2\Re\{\mathbf{q}^H\bar{\bm \phi} \}~~
        \mathrm{s.t.} ~~ \text{(\ref{eq:p1_c})},
\end{equation}
where $\mathbf{Q} \triangleq \mathbf{P}_G^H\left(\sum_{k\in\mathcal{K}}|\tau_{k}|^2\sum_{p\in\mathcal{K}}\mathbf{e}_{p,k}\mathbf{e}_{p,k}^H\right)\mathbf{P}_G$ and $\mathbf{q} \triangleq \left(\sum_{k\in\mathcal{K}}\left(\sqrt{1 + \nu_{k}} \tau_k^*\mathbf{e}_{k,k}^H-|\tau_{k}|^2\mathbf{m}_k^H\right)\mathbf{P}_G\right)^H$. 

The challenge of solving problem (\ref{eq:phi_q_phi}) is equivalent to that of solving the problem (\ref{eq:f_phi}) in the SISO case, as both are subject to the lossy BD-RIS constraints (\ref{eq:phi_F_phi_b}) – (\ref{eq:phi_F_phi_d}). The core to deal with these constraints in both problems is the ADMM framework, where the same procedure described in Section~\ref{sec:admm_method} is applied to solve (\ref{eq:phi_q_phi}). The only minor difference lies in the subproblem of variable $\bar{\bm\phi}$, which is still an unconstrained convex optimization and admits a closed-form solution given by
\begin{equation}
    \bar{\bm \phi} = \left(2\mathbf{Q}^H + \rho_1 \mathbf{C}^H\mathbf{C}\right)^{-1} \left(2\mathbf{q} + \rho_1\mathbf{C}^H(\mathbf{c} - \bm \lambda_1)\right).\label{eq:sub_phi_bar_opt} 
\end{equation}

\subsection{Complexity Analysis}
In this subsection, we provide a complexity analysis for the BCD iterative algorithm. In each iteration, the update of ${\bm \nu}$ and ${\bm \tau}$ require $\mathcal{O}\{K(M^2+MN+KN)\}$ operations, and the update of $\mathbf{W}$ has a complexity of $\mathcal{O}\{KN^2 + I_\mathrm{bs}N^3\}$, where $I_\mathrm{bs}$ is the number of iterations for the bisection search. For the update of $\bm \Phi$, the calculation of $\mathbf{Q}$ incurs a complexity of $\mathcal{O}\{K(MN+N^2) + \max(K^2\bar{M}^4, K\bar{M}M) + \bar{M}^3M^3\}$, and each iteration of updating $\bar{\bm \phi}$, $\bar{\mathbf{y}}$, and $\mathbf{z}$ involves $\mathcal{O}\{\bar{M}^3M^3\}$ operations. Hence, the overall complexity of the BCD algorithm is $\mathcal{O}\{I(I_\mathrm{bs}N^3 + KMN + \max(KM^2, K^2\bar{M}^4) + \max(K^2N, KN^2) + I'\bar{M}^3M^3)\}$, where $I$ and $I'$ denote the number of iterations for BCD and updating $\bm\Phi$, respectively. The aforementioned complexity remains unchanged when adapted for forest-connected BD-RIS.

\section{Performance Evaluation}\label{sec:simulation}
In this section, we present the simulation results to evaluate the impact of losses in the group- and forest-connected reconfigurable admittance networks on the performance of BD-RIS-aided SISO and MU-MISO systems. The channels are modeled as an integration of large-scale fading and small-scale fading. Regarding the large-scale fading, we use the distance-dependent pathloss model given by $PL_s = \zeta_0(d_s/d_0)^{-\varepsilon_s}$, $~\forall s \in \{\mathrm{RT,RI,IT}\}$. Here, $\zeta_0$ denotes the path loss at the reference distance $d_0$, $d_s$ represents the distance associated with each link ($\mathrm{RT, RI, IT}$: transmitter-to-receiver, RIS-to-receiver, and transmitter-to-RIS, respectively), and $\varepsilon_s$ corresponds to the path loss exponent for each link. For small-scale fading, the transmitter-to-receiver link is modeled as a Rayleigh fading channel, whereas the RIS-to-receiver and transmitter-to-RIS links are modeled as Rician fading channels. The Rician factors for these links are denoted by $\kappa_i$, $~\forall i \in \{\mathrm{RI,IT}\}$. In the following simulations for both SISO and MU-MISO systems, we set $\zeta_0=-30$ dB at a reference distance $d_0=1$ m for all channels. The distances are given as $d_\mathrm{RT}=52$ m, $d_\mathrm{RI}=2.5$ m and $d_\mathrm{IT}=50$ m. The path loss exponents are set to $\varepsilon_\mathrm{RT}=3.8$, $\varepsilon_\mathrm{RI}=2.2$ and $\varepsilon_\mathrm{IT}=2.5$. The Rician factors for the Rician fading channels are fixed at $\kappa_{\mathrm{RI}}=\kappa_{\mathrm{IT}}=2$ dB. The noise power is assumed as $\sigma^2=-80$ dBm for SISO systems, and $\sigma^2_k=-80$ dBm, $~\forall k\in\mathcal{K}$, for MU-MISO systems.

\begin{figure}[!t]
    \centering
    \subfigure[MM-ADMM algorithm]{
    \includegraphics[width=0.38\textwidth]{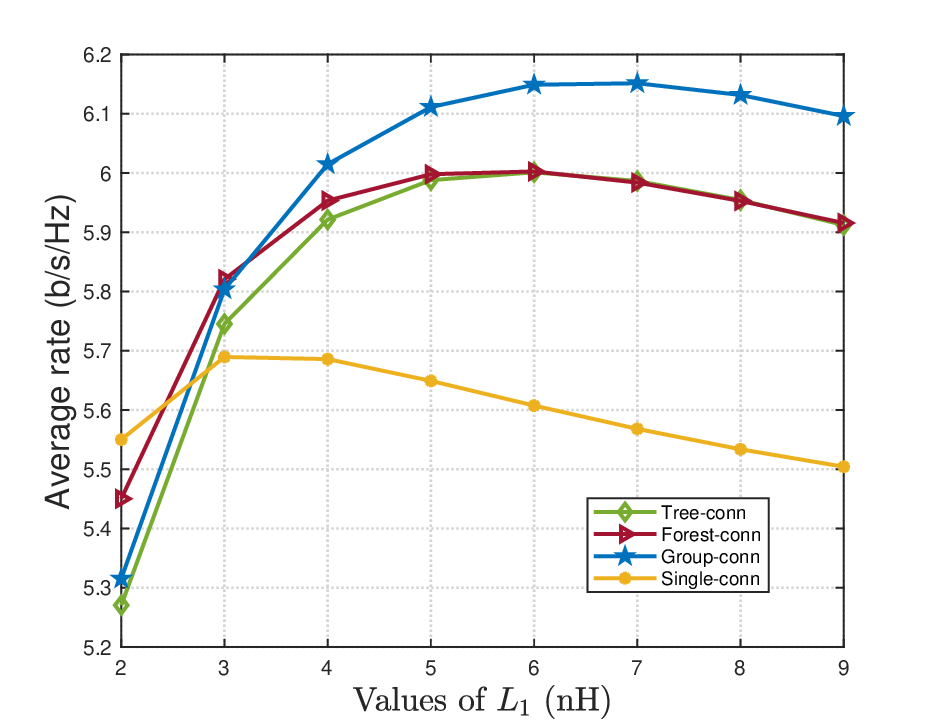}}
    \subfigure[BCD algorithm]{
    \includegraphics[width=0.38\textwidth]{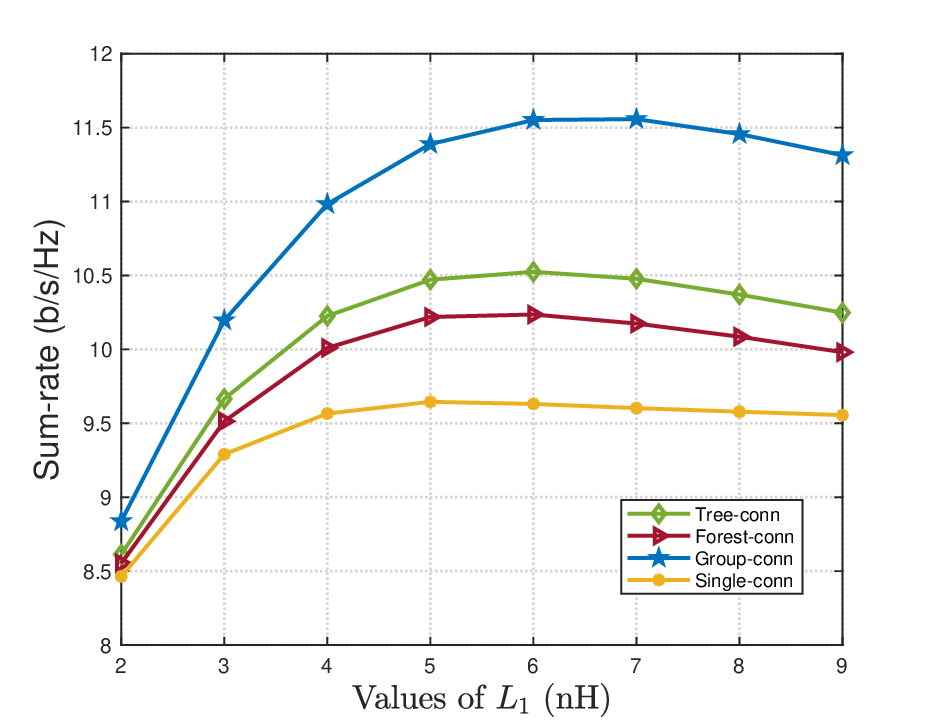}}
    \caption{(a): Average rate versus $L_1$ for SISO systems ($R=2.5\ \Omega$, $M=32$, $\bar{M}=\{1,4,32\}$, $P=20$ dBm); (b): Sum-rate versus $L_1$ for MU-MISO systems ($R=3\ \Omega$, $N=K=4$, $M=32$, $\bar{M}=\{1,4,32\}$, $P=20$ dBm).}\vspace{-0.3 cm}
    \label{fig:diff_L1}
\end{figure}

In Fig. \ref{fig:diff_L1}, we first illustrate the selection of the value for $L_1$ by plotting the average rate and sum-rate performance against different practical $L_1$ values, using MM-ADMM algorithm for SISO systems and BCD algorithm for MU-MISO systems, respectively. A practical range of parasitic inductance values is considered, and the performance of various BD-RIS architectures is evaluated within this range. For tree-connected and forest-connected architectures, the tridiagonal form is adopted. The results demonstrate that the value of $L_1$ in the circuit model has a significant impact on performance. Specifically, $L_1=6$ nH is identified as a reasonable choice, providing generally good performance in both SISO and MU-MISO systems. This value of $L_1$ is therefore used in all subsequent simulations\footnote{The values of $L_2$ and $C_{m_g,n_g},~\forall m_g,n_g$, in the simulations are determined from the datasheet of the SMV2020-079LF (see Section~\ref{sec:model_lossy_admittance}), while the value of $R$ is varied to investigate its impact on performance.}. The performance differences among BD-RIS architectures will be discussed in the next two subsections (\ref{sec:simulation_siso} and \ref{sec:simulation_miso}). From an optimization perspective, $L_1$ influences the vertical shift of the circle defined by (\ref{eq:circle}) for each tunable admittance component, thereby affecting their feasible admittance values during optimization. From a practical design perspective, a $6$ nH inductor has sufficiently low inductance to avoid introducing non-negligible parasitic resistance in the circuit \cite{Murata_LQG15WZ}.

\subsection{Convergence Behavior of the Proposed Algorithms}
\begin{figure}
    \centering
    \subfigure[Outer loop of the MM-ADMM algorithm]{
    \includegraphics[width=0.23\textwidth]{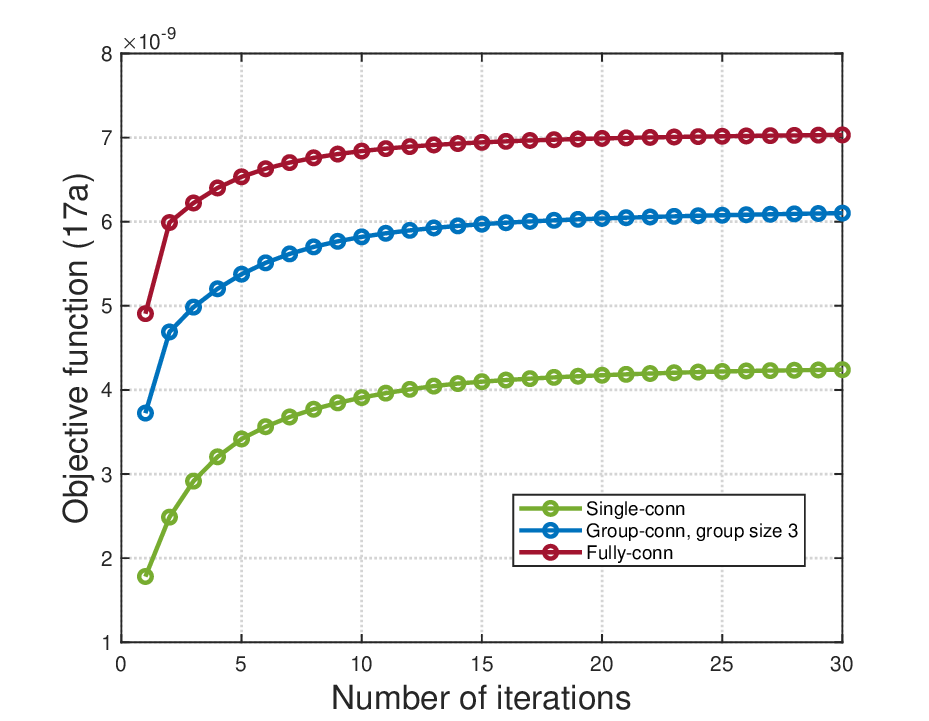}}
    \subfigure[Inner loop of the MM-ADMM algorithm]{
    \includegraphics[width=0.23\textwidth]{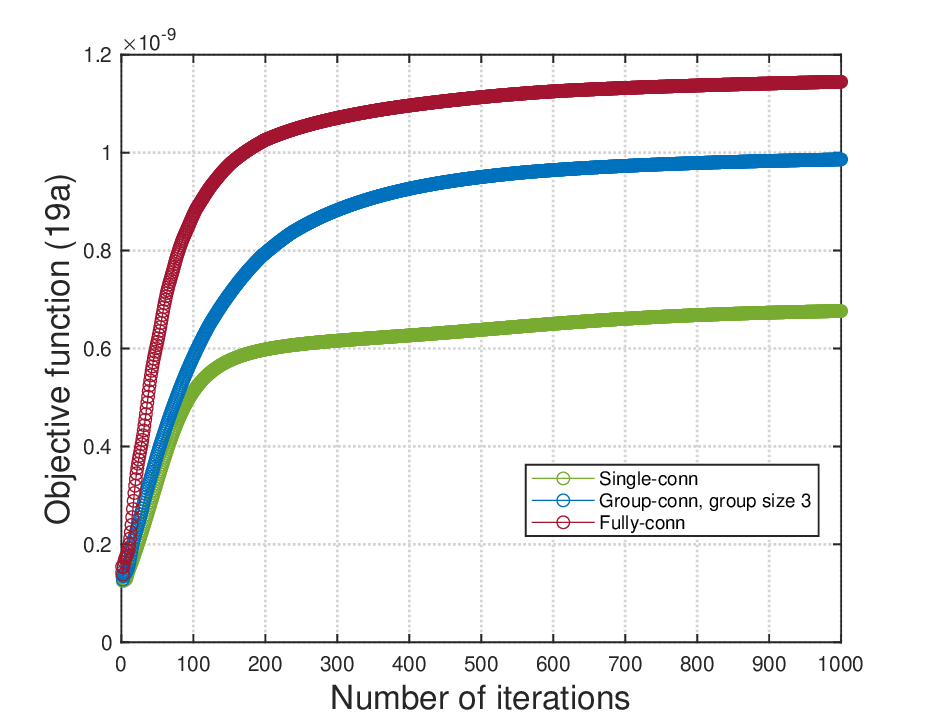}}
    \caption{(a) Convergence of the proposed MM-ADMM algorithm; (b) Convergence of the ADMM algorithm in the inner loop of MM-ADMM algorithm ($R=1 \ \Omega$, $M=30$, $\bar{M}=\{1,3,30\}$, $P=20$ dBm).}\label{fig:convergence_siso}
\end{figure}

\begin{figure}
    \centering
    \includegraphics[width=0.38\textwidth]{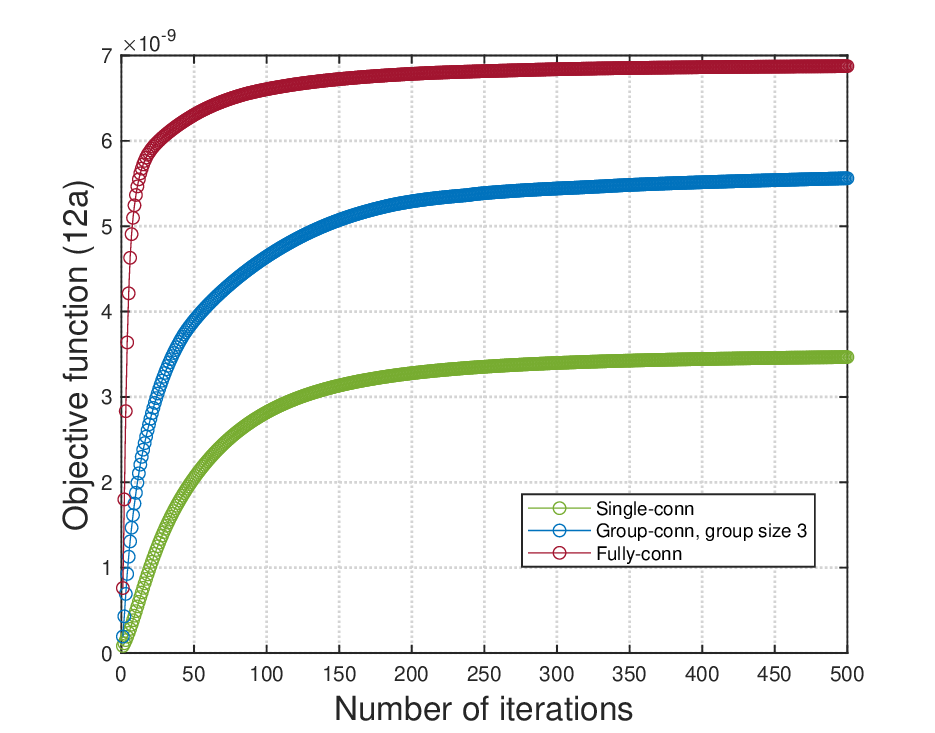}
    \caption{Convergence of the low-complexity algorithm ($R=1 \ \Omega$, $M=30$, $\bar{M}=\{1,3,30\}$, $P=20$ dBm).}
    \label{fig:convergence_low_complexity}
\end{figure}

\begin{figure}
    \centering
    \subfigure[Outer loop of the BCD algorithm]{
    \includegraphics[width=0.23\textwidth]{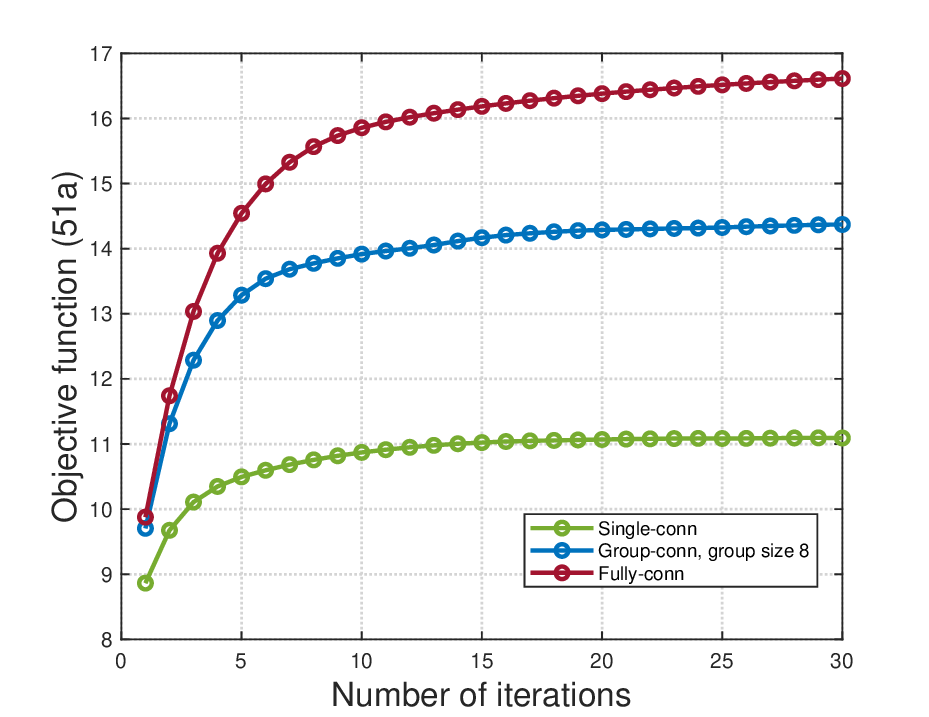}}
    \subfigure[Inner loop of the BCD algorithm]{
    \includegraphics[width=0.23\textwidth]{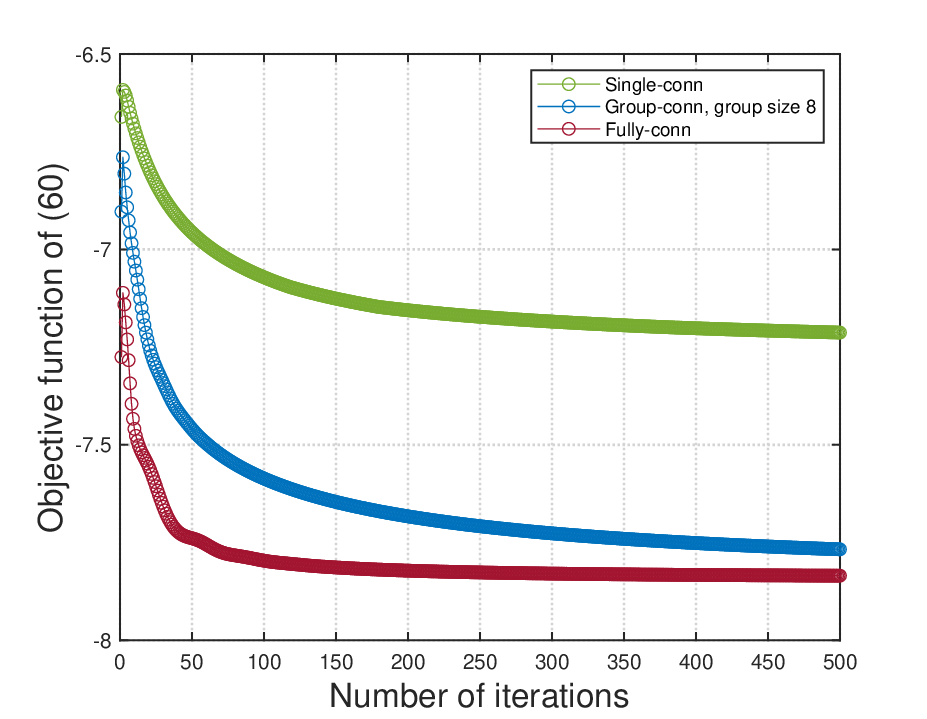}}
    \caption{(a) Convergence of the BCD algorithm; (b) Convergence of the $\bm\Phi$-subproblem update within the BCD framework ($R=1 \ \Omega$, $N=K=4$, $M=32$, $\bar{M}=\{1,8,32\}$, $P=20$ dBm).}\label{fig:convergence_mu_miso}
\end{figure}

In Figs. \ref{fig:convergence_siso} -- \ref{fig:convergence_mu_miso}, we evaluate the convergence behavior of the proposed MM-ADMM, low-complexity, and BCD algorithms, where all of these algorithms are terminated when the relative difference between two consecutive iterations is less than $10^{-4}$. As shown, all three algorithms exhibit stable convergence. In particular, the double-loop MM-ADMM and BCD algorithms converge within fewer than 30 outer iterations. Although each outer iteration requires an inner loop of about 500 iterations to update $\bm\Phi$, the small number of outer iterations ensures that the overall algorithms remain computationally efficient. Compared to MM-ADMM, the low-complexity algorithm involves only a single loop and converges more rapidly due to its much simpler objective function.

\subsection{BD-RIS-aided SISO System}\label{sec:simulation_siso}

\begin{figure*}
    \centering
    \subfigure[MM-ADMM algorithm]{
    \includegraphics[width=0.38\textwidth]{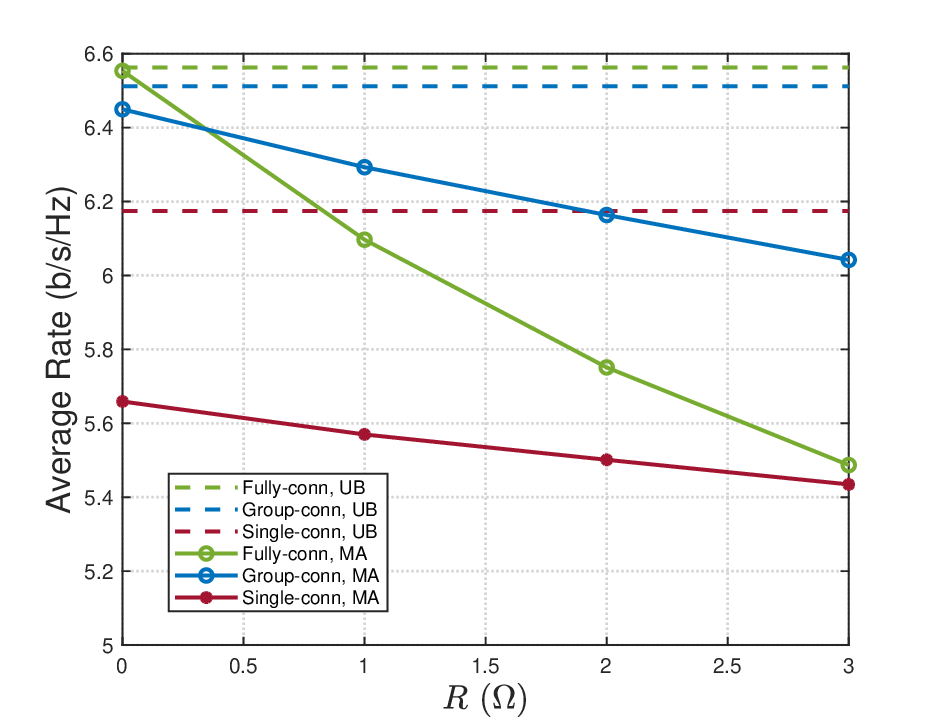}}
    \subfigure[Low-complexity algorithm]{
    \includegraphics[width=0.38\textwidth]{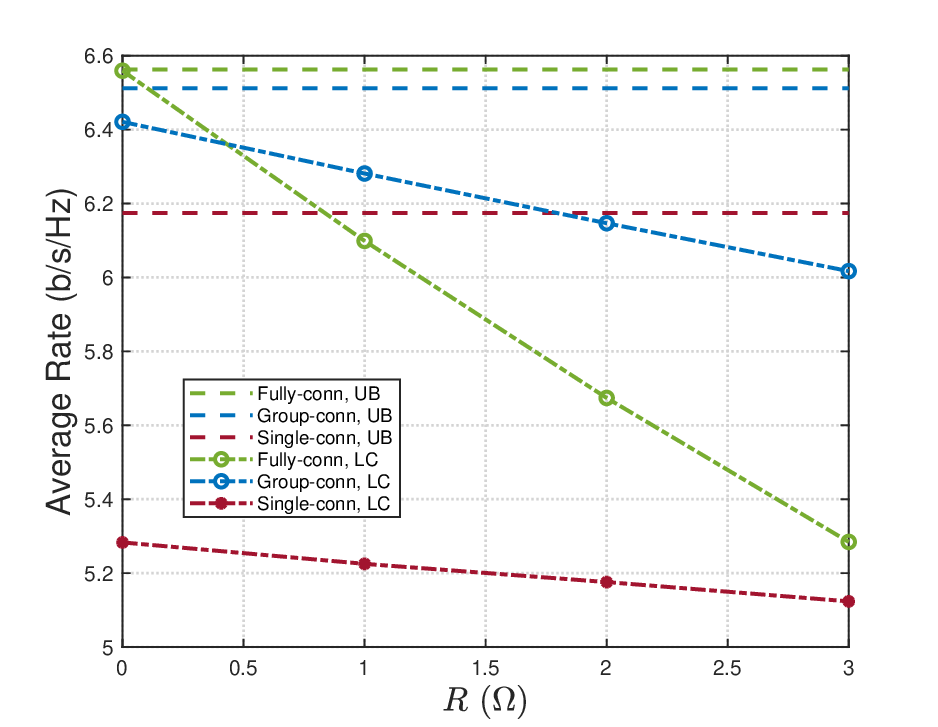}}
    \caption{Average rate versus resistance $R$ for lossy BD-RIS with group-connected architecture. In the legend, "MA" denotes MM-ADMM, "LC" denotes low-complexity, and "UB" stands for upper-bound ($M=30$, $\bar{M}\in\{1,6,30\}$, $P=20$ dBm).}
    \label{fig:average_rate_R_group}
\end{figure*}

\begin{figure*}
    \centering
    \subfigure[MM-ADMM algorithm]{
    \includegraphics[width=0.38\textwidth]{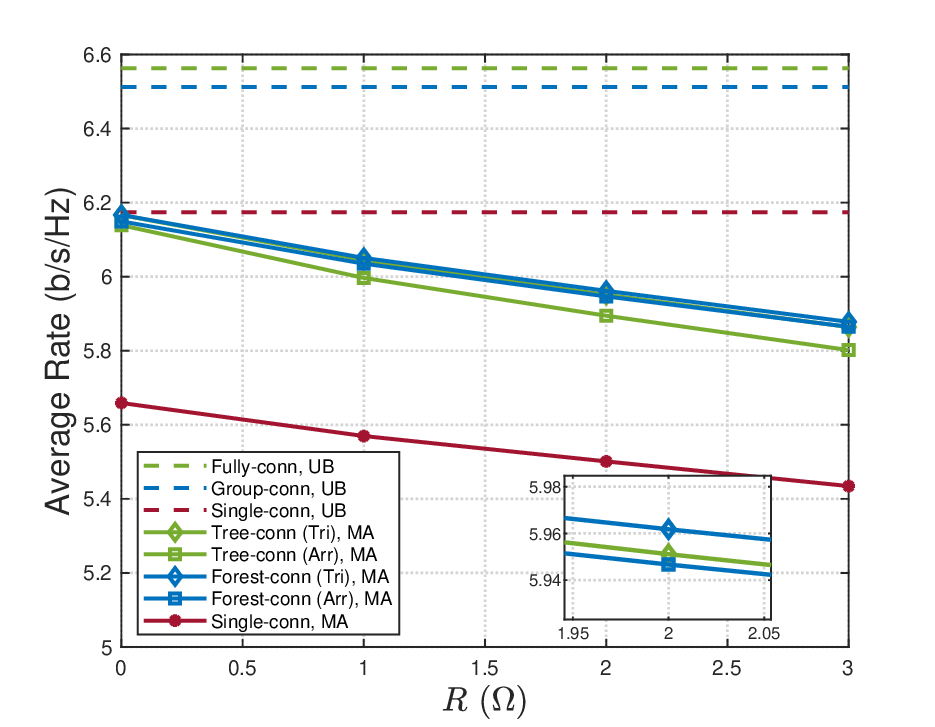}}
    \subfigure[Low-complexity algorithm]{
    \includegraphics[width=0.38\textwidth]{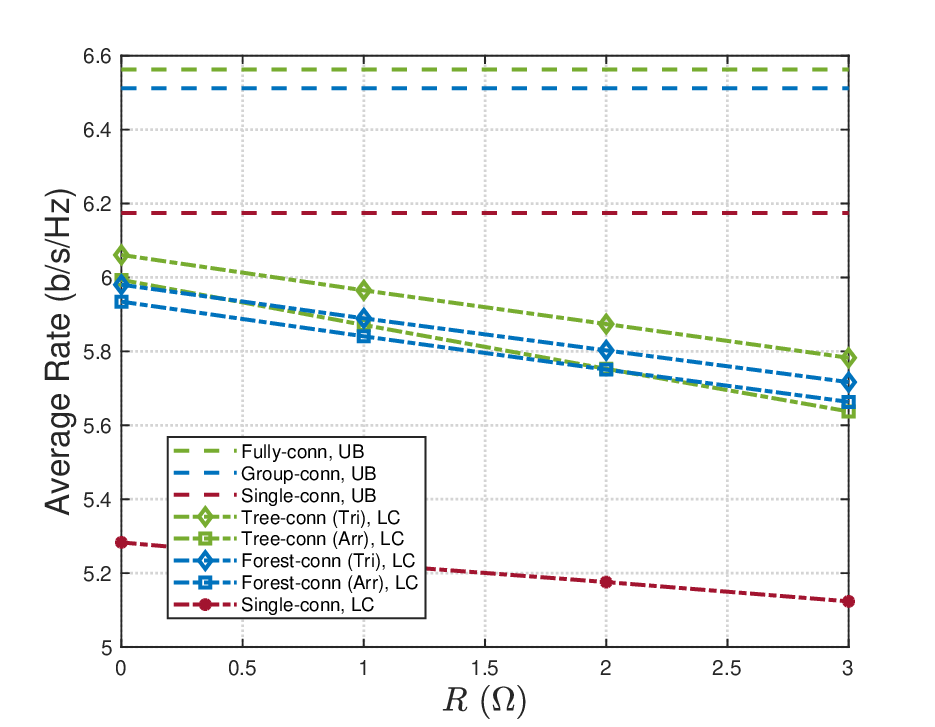}}
    \caption{Average rate versus resistance $R$ for lossy BD-RIS with forest-connected architectures. In the legend, "MA" denotes MM-ADMM, "LC" denotes low-complexity, "Tri" stands for tridiagonal, "Arr" stands for arrowhead, and "UB" stands for upper-bound ($M=30$, $\bar{M}\in\{1,6,30\}$, $P=20$ dBm).}
    \label{fig:average_rate_R_forest}
\end{figure*}

We first evaluate the performance of the proposed MM-ADMM algorithm and the low-complexity algorithm for BD-RIS-aided SISO systems by plotting the average rate versus resistance $R$ in Figs. \ref{fig:average_rate_R_group} and \ref{fig:average_rate_R_forest}. The rate for each channel realization is computed as $\log_2(1 + P\frac{|h|^2}{\sigma^2})$, and the upperbound\footnote{The gap between the upperbound and the rate at $R=0 \ \Omega$ arises because, at $R=0 \ \Omega$, each tunable admittance remains constrained to the imaginary axis with bounded capacitance values $C_{m_g,n_g}$.} for the maximum received signal power of group- and forest-connected BD-RISs is given by $\left(\sum_{g=1}^G \|\mathbf{h}_{RI,g} \| \|\mathbf{h}_{IT,g}\| + |h_{RT}|\right)^2$. We plot the performance of both algorithms for group-connected BD-RIS in Fig. \ref{fig:average_rate_R_group}, from which we have the following observations.

\textit{First}, for both algorithms, BD-RIS with a group-connected architecture outperforms D-RIS across the entire considered range of $R$. With a relatively small group size, e.g., $\bar{M}=6$, group-connected BD-RIS maintains superior performance over D-RIS, exhibiting only marginally faster degradation while delivering substantially better performance. This is because when the circuit complexity is relatively low, the enhanced wave manipulation flexibility provided by more reconfigurable admittance components of BD-RIS outweighs their losses. However, this does not hold for fully-connected cases since the quadratically increased admittances introduce significant losses. As a result, the fully-connected BD-RIS with $\bar{M}=M$ exhibits significantly faster performance degradation compared to D-RIS. This indicates that the losses in BD-RIS have the most significant impact on the fully-connected architecture, implying that the group-connected architecture with a proper group size is more practical. 

\textit{Second}, although the MM-ADMM algorithm generally outperforms the low-complexity algorithm for D-RIS, their performance is nearly identical for fully-connected BD-RIS when $R\leq 2\ \Omega$ and group-connected BD-RIS with $\bar{M}=6$. For D-RIS, the average rate achieved by MM–ADMM exceeds that of the low-complexity solution by 7.12\%, 6.60\%, 6.29\%, and 6.07\% for $R = \{0,1,2,3\}\ \Omega$, respectively. For fully-connected case when $R=3\ \Omega$, the performance gap is 3.83\%. These observations indicate that the approximation in the low-complexity algorithm becomes less accurate for D-RIS and for high-loss fully-connected BD-RIS.

In Fig. \ref{fig:average_rate_R_forest}, the performance of both algorithms for forest-connected BD-RIS is presented, from which the following observations can be made.

\textit{First}, for both algorithms, forest-connected BD-RIS with both tridiagonal and arrowhead forms outperform D-RIS across the entire considered range of $R$. Both algorithms show no significant performance difference between forest-connected BD-RIS with $\bar{M} = 6$ and tree-connected BD-RIS with $\bar{M} = M$. This is because the tree-connected architecture introduces only a small number of additional tunable admittances compared to the forest-connected one. For the low-complexity algorithm, these extra tunable admittances help to compensate for approximation errors. As a result, the tridiagonal form of the tree-connected BD-RIS outperforms the forest-connected BD-RIS under the low-complexity algorithm. 

\textit{Second}, the performance gap between the low-complexity and MM-ADMM algorithm remains relatively small for all forest- and tree-connected architectures. For example, for the tridiagonal tree-connected BD-RIS, MM-ADMM outperforms the low-complexity solution by only 1.71\%, 1.33\%, 1.31\%, and 1.41\% for $R = \{0,1,2,3\}\ \Omega$, respectively, indicating that the low-complexity algorithm is well suited to forest-connected architectures given its much lower computational complexity.

In both Figs. \ref{fig:average_rate_R_group} and \ref{fig:average_rate_R_forest}, when accounting for lossy tunable admittances, both algorithms demonstrate that group-connected BD-RIS (excluding the fully-connected architecture) outperforms forest-connected BD-RIS. This observation is not consistent with the conclusion for lossless cases, where the group- and forest-connected BD-RISs could achieve the same performance \cite{nerinigraphtheory}. This is because, lossless admittances have purely imaginary and arbitrarily tunable admittance values, where $(2\bar{M}-1)G$ components are sufficient to reach the performance upperbound. However, in lossy cases, both architectures cannot reach the performance upperbound, while the former has more components to be tuned to compensate for the power loss.

\begin{figure}[!t]
    \centering
    \includegraphics[width=0.38\textwidth]{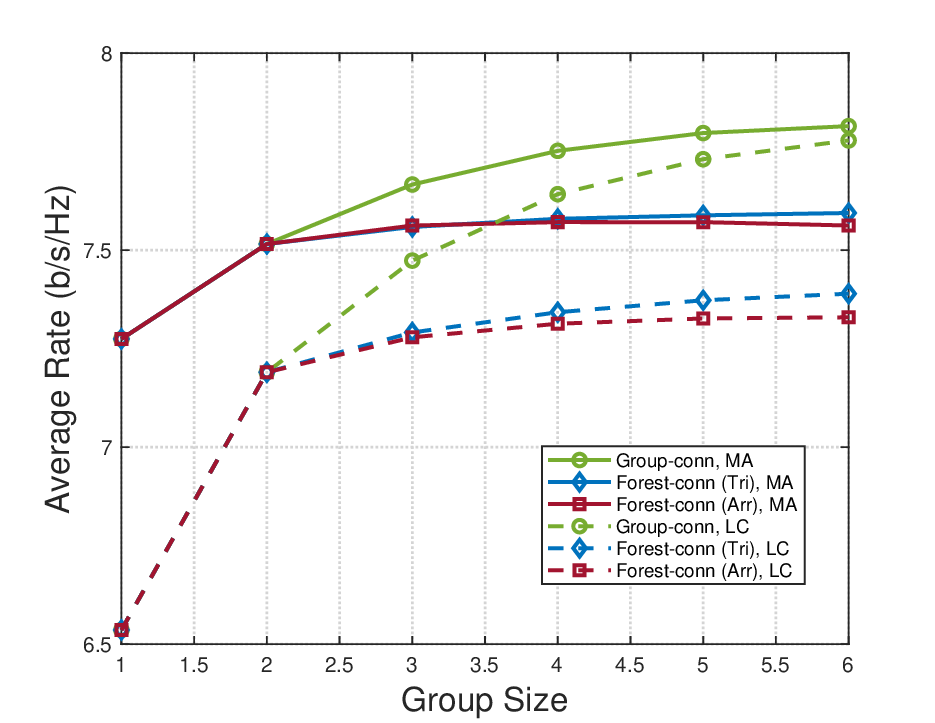}
    \caption{Average rate versus group size $\bar{M}$ for lossy BD-RIS with group- and forest-connected architectures. In the legend, "MA" denotes MM-ADMM, "LC" denotes low-complexity, "Tri" stands for tridiagonal, "Arr" stands for arrowhead, and "UB" stands for upper-bound ($M=60$, $P=20$ dBm, $R=2.5 \ \Omega$).}\vspace{-0.3 cm}
    \label{fig:average_rate_MG}
\end{figure}

Fig. \ref{fig:average_rate_MG} compares the performance of group-connected and forest-connected BD-RISs as the group size $\bar{M}$ increases, evaluating the MM-ADMM and low-complexity algorithms. The results reveal two interesting trends: First, for group-connected BD-RIS, the performance gap between two algorithms narrows as $\bar{M}$ increases, demonstrating that a larger group size provides greater flexibility to compensate for approximation errors. Second, for both algorithms, forest-connected BD-RIS exhibits only marginal performance improvements as $\bar{M}$ increases from 3 to 6. This occurs because the increase in $\bar{M}$ within $\{3,\ldots,6\}$ introduces only a small number of additional tunable admittances. In summary, a relative small group size (e.g., $\bar{M}\in\{4,5,6\}$) in the group-connected architecture with the low-complexity algorithm can achieve good performance while keeping computational complexity low.

\subsection{BD-RIS-aided MU-MISO System}\label{sec:simulation_miso}

\begin{figure}[!t]
    \centering
    \includegraphics[width=0.38\textwidth]{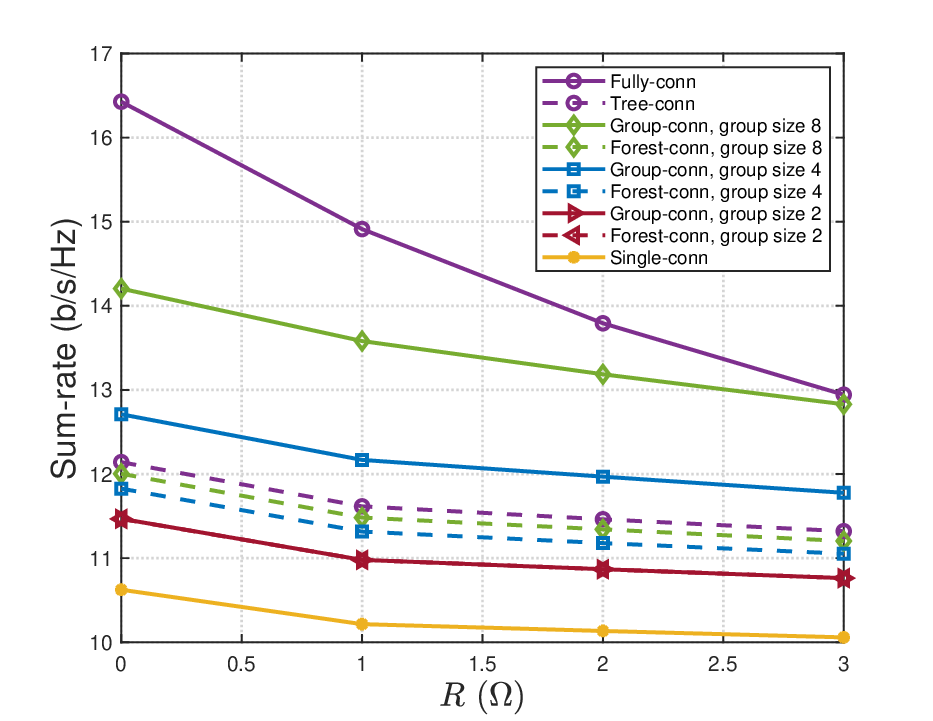}
    \caption{Sum-rate versus resistance $R$ for lossy BD-RIS with group- and forest-connected architectures ($N=K=4$, $M=32$, $\bar{M}\in\{1,2,4,8,32\}$, $P=20$ dBm).}\vspace{-0.3 cm}
    \label{fig:sum_rate_R}
\end{figure}

We then evaluate the sum-rate performance of the BD-RIS-aided MU-MISO system by plotting the sum-rate versus resistance $R$ in Fig. \ref{fig:sum_rate_R}. The simulations employ the tridiagonal form of the forest-connected architecture, with both the number of users, $K$, and the number of antennas at the BS, $N$, fixed to 4. From Fig. \ref{fig:sum_rate_R}, we have the following observations.

\textit{First}, all BD-RIS architectures outperform D-RIS across the entire considered range of $R$. Unlike the SISO case where fully-connected BD-RIS achieves comparable performance to D-RIS at $R=3\ \Omega$, in MU-MISO case, fully-connected BD-RIS maintains significant gains over D-RIS and outperforms other architectures even under a large $R$. This enhanced performance likely stems from the proper precoder design effectively compensating for the huge losses at BD-RIS.

\textit{Second}, BD-RIS with group-connected architectures outperforms forest-connected architectures under the same group size. The sum-rate improvement achieved by increasing $\bar{M}$ is substantially greater for group-connected architectures than for forest-connected architectures, which is consistent with the observations in SISO systems. This is because each $\bar{M}$ increment in group-connected BD-RIS introduces more tunable admittances than in forest-connected BD-RIS. These results suggest that a larger number of tunable admittances helps to combat losses.

\begin{figure}[!t]
    \centering
    \includegraphics[width=0.38\textwidth]{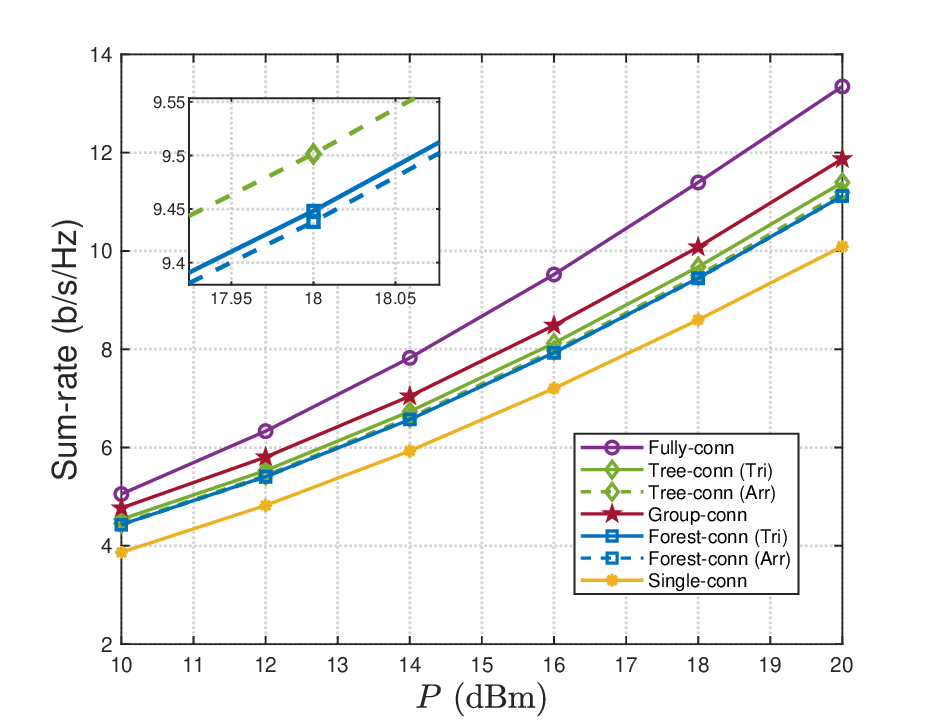}
    \caption{Sum-rate versus transmit power $P$ for lossy BD-RIS with group- and forest-connected architectures. In the legend, "Tri" stands for tridiagonal, and "Arr" stands for arrowhead ($N=K=4$, $M=32$, $\bar{M}\in\{1,4,32\}$, $R=2.5\ \Omega$).}
    \label{fig:sum_rate_P}
\end{figure}

In Fig. \ref{fig:sum_rate_P}, we compare the sum-rate performance of group- and forest-connected BD-RISs versus the transmit power $P$. As expected, fully-connected BD-RIS achieves the best performance, and the group-connected BD-RIS outperforms forest-connected BD-RIS with the same group size. The presence of more interconnections in lossy BD-RIS is beneficial in MU-MISO systems compared to single-user SISO systems. This is because the multiuser MISO channel provides a higher degree of freedom (DoF) and relies more heavily on interconnections to exploit this increased DoF \cite{band_stem_zheyu}. As a result, additional tunable admittances enhance performance, even in the presence of losses.

\begin{figure}[!t]
    \centering
    \includegraphics[width=0.38\textwidth]{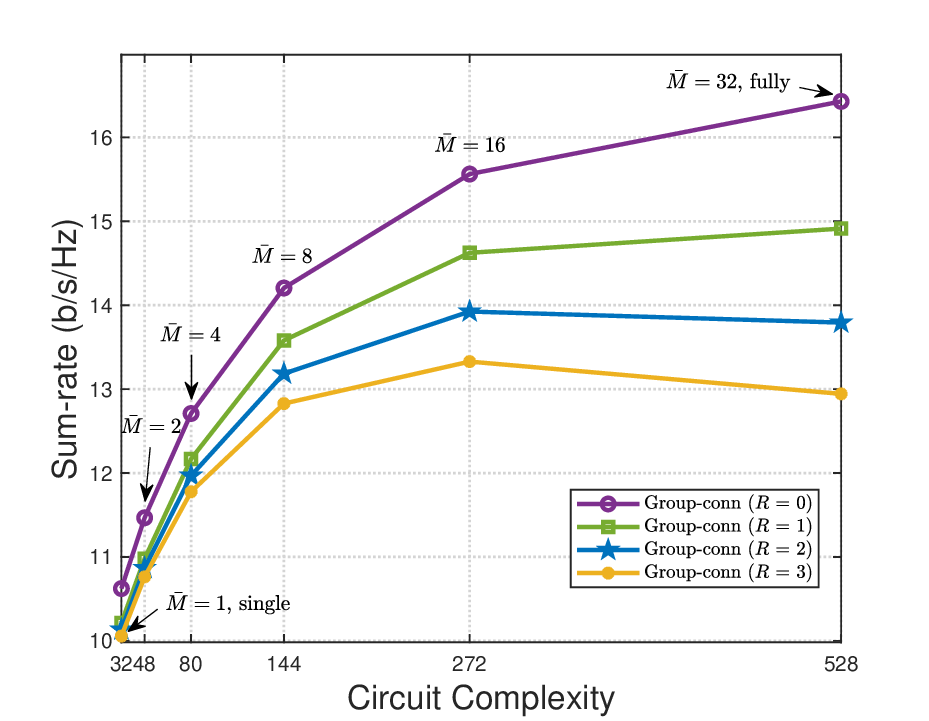}
    \caption{Performance-complexity trade-off for lossy BD-RIS with a group-connected architecture, given corresponding group size $\bar{M}$ and different values of $R$ ($N=K=4$, $M=32$, $P=20$ dBm).}
    \label{fig:sum_rate_complexity}
\end{figure}

In Fig. \ref{fig:sum_rate_complexity}, we plot the sum-rate performance versus circuit complexity for the group-connected BD-RIS. Unlike the Pareto frontier of the lossless BD-RIS reported in \cite{band_stem_zheyu}, increasing the circuit complexity does not always lead to higher performance; for example, the fully-connected architecture fails to achieve the best performance when $R\geq 2\ \Omega$. Under significant losses (i.e., under large $R$), adding more tunable components may in fact degrade the overall performance. This observation highlights the importance of carefully balancing performance and complexity, emphasizing the need to select an appropriate level of complexity for lossy BD-RIS.

\section{Conclusion}\label{sec:conclusion}
In this paper, we investigate lossy BD-RIS modeling and optimization for SISO and MU-MISO systems. Specifically, we model the lossy reconfigurable admittance network by representing each tunable admittance component with a practical varactor diode based on a lumped circuit model. Taking into account the practical capacitance range of the varactor, each tunable admittance is constrained within a feasible range.

Building on the proposed lossy BD-RIS model, we optimize the scattering matrix for BD-RIS-aided SISO systems to maximize the received signal power. To solve this problem, we employ an MM framework to transform it into a more tractable form and then develop an ADMM algorithm to efficiently handle the complex constraints. To reduce computational demands, we further propose a low-complexity algorithm that approximates the original problem. Next, we extend the optimization to the joint design of the transmit precoder and scattering matrix to maximize the sum-rate for BD-RIS-aided MU-MISO systems. This problem is reformulated as a multi-block optimization using FP theory, and an ADMM algorithm, similar to the SISO case, is applied to efficiently update the scattering matrix.

Under the proposed lossy BD-RIS model, simulations results show that all BD-RIS architectures outperform D-RIS in both SISO and MU-MISO systems. However, the optimal BD-RIS architectures in the lossless case are not necessarily optimal when losses are considered, e.g. fully-connected and tree-connected BD-RISs fail to achieve the best performance in SISO systems. This suggests that the characterization of the fundamental Pareto frontier between performance and complexity for lossless architectures in \cite{parteo_frontier_nerini,band_stem_zheyu} needs to be revisited in the lossy case. A thorough investigation of the performance–complexity trade-off for lossy BD-RIS architectures is thus left as an insightful future work.

\begin{appendix}[Explicit forms of $\mathbf{P}$ and $U$]
\textit{i}) Group-connected architecture: $U=\frac{\bar{M}(\bar{M}+1)}{2}$ and $\bar{\mathbf{y}}_g$ collects the $\bar{M}$ diagonal and $\frac{\bar{M}(\bar{M}-1)}{2}$ upper triangular entries of $\bar{\mathbf{Y}}_g$. The mapping matrix $\mathbf{P}\in \{0,1\}^{\bar{M}^2 \times \frac{\bar{M}(\bar{M}+1)}{2}}$ is given by\begin{equation}\label{eq:P_group}
    \begin{aligned}
        \relax[\mathbf{P}]_{\bar{M}(i-1)+j,k} = \qquad\qquad\qquad\qquad\qquad\qquad\qquad\qquad\\
        \begin{cases}
            1, & k = \frac{(2\bar{M}-j)(j-1)}{2} + i ~\text{and}~ 1 \le j \le i;\\
            1, & k = \frac{(2\bar{M}-i)(i-1)}{2} + j ~\text{and}~ i<j\le \bar{M};\\
            0, & \text{otherwise},
        \end{cases}
    \end{aligned}
\end{equation}$\forall i,j \in \mathcal{\bar{M}}$.

\textit{ii}) Forest-connected architecture (tridiagonal form): $U = 2\bar{M}-1$ and $\bar{\mathbf{y}}_g$ collects the $\bar{M}$ diagonal entries and $\bar{M}-1$ off-diagonal entries of $\bar{\mathbf{Y}}_g$ based on (\ref{eq:Y_tri}). The mapping matrix $\mathbf{P}\in \{0,1\}^{\bar{M}^2 \times (2\bar{M}-1)}$ is given by\begin{equation}\label{eq:P_tri}
     \begin{aligned}
        \relax[\mathbf{P}]_{\bar{M}(i-1)+j,k} = \qquad\qquad\qquad\qquad\qquad\qquad\qquad\qquad\quad\\
        \begin{cases}
            1, & k = j, ~i=1, ~\text{and}~  j \le 2;\\
            1, & k = \bar{M}+j-1,~i=\bar{M} ~\text{and}~ j > i-1;\\
            1, & k = i-1+j,~1<i<\bar{M} ~\text{and}~ i-1\leq j \leq i+1;\\
            0, & \text{otherwise},
        \end{cases}
    \end{aligned}
\end{equation}
$\forall i,j \in \mathcal{\bar{M}}$.

\textit{iii}) Forest-connected architecture (arrowhead form): $U = 2\bar{M}-1$ and $\bar{\mathbf{y}}_g$ collects the $\bar{M}$ diagonal entries and $\bar{M}-1$ off-diagonal entries of $\bar{\mathbf{Y}}_g$ based on (\ref{eq:Y_arr}). The mapping matrix $\mathbf{P}\in \{0,1\}^{\bar{M}^2 \times (2\bar{M}-1)}$ is given by\begin{equation}\label{eq:P_arr}
     \begin{aligned}
        \relax[\mathbf{P}]_{\bar{M}(i-1)+j,k} =
        \begin{cases}
            1, & k = j, ~\text{and}~  i = 1;\\
            1, & k = i,~j=1 ~\text{and}~ i > 1;\\
            1, & k = \bar{M}+i-1,~i=i ~\text{and}~ i>1;\\
            0, & \text{otherwise},
        \end{cases}
    \end{aligned}
\end{equation}
$\forall i,j \in \mathcal{\bar{M}}$.
\end{appendix}

\bibliographystyle{IEEEtran}
\bibliography{refs}

\end{document}